\documentclass[amsmath,amssymb,aps,twocolumn,floats,prx,unsortedaddress,superscriptaddress]{revtex4-1}
\usepackage{hyperref} 
\usepackage{graphicx}
\usepackage{textcomp}
\usepackage{amsmath}
\usepackage{amssymb}
\usepackage{color}
\usepackage{ulem}
\usepackage[mathscr]{euscript}
\usepackage[margin=1in]{geometry}
\usepackage{verbatim}
\usepackage{slashed}
\usepackage[utf8]{inputenc}
\usepackage{floatrow}

\newcommand{\bfig}{\begin{figure}}
\newcommand{\efig}{\end{figure}}

\allowdisplaybreaks

\usepackage{float}

\newcommand{\bea}{\begin{eqnarray}}
\newcommand{\ena}{\end{eqnarray}}
\newcommand{\bee}{\begin{equation}}
\newcommand{\ene}{\end{equation}}

\newcommand{\res}{\mathrm{res}}
\newcommand{\field}{\mu_0H}

\newcommand{\delk}{\Delta\kappa}
\newcommand{\kxx}{\kappa}
\newcommand{\wxx}{\kappa^{-1}}

\newcommand{\kph}{\kappa_{\mathrm{ph}}}
\newcommand{\wph}{\kappa_{\mathrm{ph}}^{-1}}
\newcommand{\kmag}{\kappa_{\mathrm{mag}}}

\newcommand{\nmag}{n_{\mathrm{mag}}}

\newcommand{\cmag}{c_{\mathrm{mag}}}
\newcommand{\cph}{c_{\mathrm{ph}}}

\newcommand{\ka}{\kappa}
\newcommand{\rucl} {RuCl$_{3}$}
\newcommand{\crcl} {CrCl$_{3}$}

\newcommand{\sccl} {ScCl$_{3}$}


\begin{document}



\title{Giant thermal magnetoconductivity in \crcl~and a general model for 
spin-phonon scattering}


\author{Christopher A. Pocs}
\affiliation{Department of Physics, University of Colorado, Boulder, CO 80309, 
USA}%
\author{Ian A. Leahy}
\affiliation{Department of Physics, University of Colorado, Boulder, CO 80309, 
USA}%
\author {Hao Zheng}
\affiliation{Department of Physics, University of Colorado, Boulder, CO 80309, 
USA}%
\author{Gang Cao}
\affiliation{Department of Physics, University of Colorado, Boulder, CO 80309, 
USA}%
\author{Eun-Sang Choi}
\affiliation{National High Magnetic Field Laboratory, 1800 E. Paul Dirac Dr.,
Tallahassee, FL 32310-3706, USA}%
\author{S.-H. Do}
\affiliation{Department of Physics, Chung-Ang University, Seoul, 790-784, 
South Korea}%
\author{Kwang-Yong Choi}
\affiliation{Department of Physics, Chung-Ang University, Seoul, 790-784, 
South Korea}%
\author{B. Normand}
\affiliation{Neutrons and Muons Research Division, Paul Scherrer Institute, 
CH-5232 Villigen PSI, Switzerland}%
\author{Minhyea Lee}
\affiliation{Department of Physics, University of Colorado, Boulder, CO 80309, 
USA}%

\date{\today}

\begin{abstract}
Insulating quantum magnets lie at the forefront both of fundamental research 
into quantum matter and of technological exploitation in the increasingly 
applied field of spintronics. In this context, the magnetic thermal transport 
is a particularly sensitive probe of the elementary spin and exotic topological 
excitations in unconventional magnetic materials. However, magnetic 
contributions to heat conduction are invariably intertwined with lattice 
(phonon) contributions, and thus the issue of spin-phonon coupling in 
determining the spin and thermal transport properties of magnetic insulators 
becomes more important with every emergent topological magnetic system.
Here we report the observation of an anomalously strong enhancement of the 
thermal conductivity, occurring at all relevant temperatures, in the layered 
honeycomb material \crcl~in the presence of an applied magnetic field. Away 
from the magnetically ordered phase at low temperatures and small 
fields, there is no coherent spin contribution to the thermal conductivity, 
and hence the effect must be caused by a strong suppression of the phonon 
thermal conductivity due to magnetic fluctations, which are in turn suppressed 
by the field. We build an empirical model for the thermal conductivity of 
\crcl~within a formalism assuming an independently determined number of 
spin-flip processes and an efficiency of the phonon scattering events they 
mediate. By extracting the intrinsic phonon thermal conductivity we obtain a 
quantitative description of our measured data at all fields and temperatures, 
and we demonstrate that the scattering efficiency is entirely independent of 
the applied field. In this way we use \crcl~as a model system to understand 
the interactions between spin and phonon excitations in the context of thermal 
transport. We anticipate that the completely general framework we introduce 
will have broad implications for the interpretation of transport phenomena 
in magnetic quantum materials. 
\end{abstract}

\maketitle

\section{Introduction}
\label{sintro}

Transport measurements form one of the three pillars of experimental 
condensed matter physics. In insulating crystalline systems the thermal 
conductivity, $\ka(T)$, ranks as one of the most valuable probes for 
investigating the low-energy excitations \cite{Zimanbook}. Unlike thermal 
equilibrium quantities such as the specific heat, $c(T)$, $\ka(T)$ is a 
steady-state transport property and thus contains fundamental information 
about the itinerant characteristics of a system, most notably the relaxation 
times and scattering strengths of the low-energy excitations. In the field of 
insulating quantum magnets, low-dimensional spin systems may host a wide range 
of exotic ground states and $\ka(T)$ has long been one of the most important 
probes of their unconventional spin excitations \cite{Ando1998,Sologubenko2000,
Hofmann2001,Sales2002,Jin2003,SYLi2005,Wu2015,Leahy2017}. Because the lattice 
phonons invariably constitute a strong and relatively well-characterized 
contribution to thermal transport, $\ka(T)$ measurements offer particular 
insight into the dominant spin-phonon scattering mechanisms. Even in materials 
whose thermal conductivity is phonon-dominated, meaning that there is no 
significant heat transport due to coherent magnetic modes, a strong 
field-dependence of $\ka(T)$ may still be present due to destructive 
effects of the spin sector on the phonon transport. 

\crcl~is an insulating, layered, honeycomb-lattice compound and has attracted 
considerable recent attention from two independent lines of research. One 
concerns the ``candidate Kitaev'' material $\alpha$-RuCl$_3$ \cite{rfgft,rpea,
rsea,rjea,rcaoetal}, whose proximity to Kitaev physics may be gauged from the 
nature of its magnetic excitations \cite{Banerjee2016,Zheng2017,Ponomaryov2017,
Banerjee2017,Banerjee2018}. While $\ka(T)$ measurements have not been able 
to provide conclusive evidence of fractionalized fermionic spin modes in 
$\alpha$-RuCl$_3$, one of the primary reasons why the issue remains open 
concerns the role and indeed the nature of spin-phonon scattering, which 
is manifestly strong over a wide range of temperatures \cite{Leahy2017,
Hentrich2018}. \crcl~is the 3d transition-metal structural analog of 
$\alpha$-RuCl$_3$, and as such represents the latter material in the absence 
of significant spin-orbit coupling. Although one may fear that this removal 
of Kitaev character removes any connection between the two systems, we will 
show here that \crcl~presents a test case for spin-phonon scattering effects 
that are at least as strong as any comparable phenomena in $\alpha$-RuCl$_3$. 

The second avenue leading to \crcl~as a key material to understand is its 
position in the structural series CrX$_3$, where X $=$ Cl, Br, I is a halide. 
Structurally, the chromium trihalides are van der Waals materials, allowing 
them to be cleaved easily and prepared in mono- or few-layer forms that show 
strong differences in their physical properties. Magnetically, the honeycomb 
layers have ferromagnetic (FM) in-plane interactions and indeed both CrI$_3$ 
and CrBr$_3$ are bulk ferromagnets. This magnetic character has therefore 
promoted a keen interest in spin and lattice control in the context of 
topological magnonics \cite{Pershoguba2018}, spintronics, and 
magnetoelectronics \cite{rmdcs,Huang2017}. \crcl~is a historical ``mixed 
FM/AF'' system, with antiferromagnetic (AF) interlayer interactions ensuring 
an AF ground state of anti-aligned FM layers \cite{Hansen1958,Cable1961,
Bizette1961,Narath1965,Kuhlow1982}, and because of the rather low field 
scale ($\mu_0 H_s \simeq 2$ T, independent of direction) for complete spin 
polarization is an excellent candidate for studying magnetic-field effects 
on spin and lattice transport. 

\begin{figure}[t]
\includegraphics[width=\linewidth]{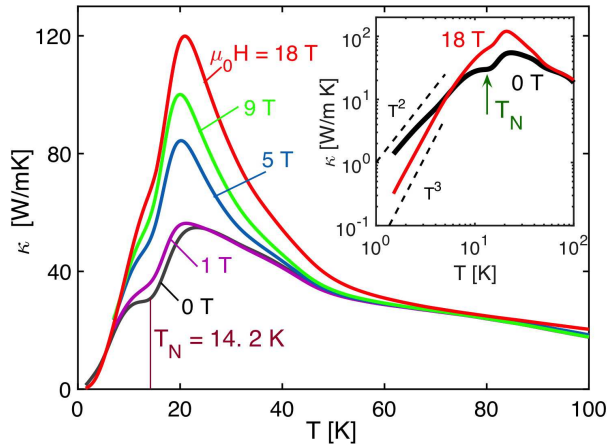}
\caption{$\kxx$ shown as a function of $T$ for different values of the magnetic 
field applied parallel to the temperature gradient in the $ab$ plane. Inset: 
$\kxx(T)$ shown on a logarithmic scale for comparison of the low-$T$ data with 
the power-law forms $\ka(T) \propto T^2$ and $T^3$.}
\label{fig:kT}
\end{figure}

In Fig.~\ref{fig:kT} we show our key experimental result, that the in-plane 
thermal conductivity of \crcl~is greatly enhanced by a magnetic field. We 
comment for clarity that by ``in-plane'' we refer to measurements of $\ka(T)$ 
performed with the temperature gradient, $\nabla T$, oriented in the $ab$-plane 
and with the field also applied in the plane ($H \parallel \nabla T$). 
Qualitatively, this field-dependence of $\kxx(T)$ resembles the behavior 
typical of magnetoresistance in magnetic conductors, where the electrical 
conductivity often increases as the field is increased \cite{rgmr}; this 
effect is caused by the field-induced reduction of the spin-dependent 
scattering and is often most prominent in the vicinity of the magnetic 
ordering transition. In \crcl, the origin of this giant thermal 
magnetoresistance lies in the very strong field-induced suppression of the 
spin-phonon scattering, and one may also observe in Fig.~\ref{fig:kT} that it 
is particularly prominent at the ordering transition ($T_N$). Quantitatively, 
the phenomenon is anomalously large and is particularly unusual in that it 
extends over essentially the entire range of temperatures shown in 
Fig.~\ref{fig:kT}. Here we will show that it can be modelled accurately 
with a minimal number of assumptions and empirical parameters. 

In many cases, the phonon thermal conductivity is well captured by the 
highly refined Debye-Callaway (DC) model \cite{Callaway1959,Bergmanbook}, 
in which all of the most important mechanisms for phonon scattering are 
considered over a wide range of temperatures. One obtains the form 
\bea
\kph = \frac{k_B^4}{2\pi^2 v_s \hbar^3} \, T^3 \!\!\! \int_{0}^{\Theta_D/T} 
\!\!\!\!\! \frac{x^4 \, e^x}{(e^x - 1)^2} \, \tau(\omega, T) \, dx,
\label{eq:DC}
\ena
where $\omega$ is the phonon frequency, $\Theta_D$ is the Debye temperature, 
$v_s$ is a characteristic average phonon velocity, and the integration variable 
is $x = \hbar \omega/k_B T$. In the relaxation-time approximation, it is 
assumed that all possible scattering mechanisms contribute independently to 
the phonon scattering time, $\tau (\omega,T)$, and hence 
\bee
\tau^{-1} = \tau_b^{-1} + \tau_{pd}^{-1} + \tau_{U}^{-1} + \tau_{\res}^{-1}, 
\label{eq:tau}
\ene
where the four relaxation times account respectively for boundary scattering, 
point-defect scattering, Umklapp scattering, and resonant scattering due 
to impurities, magnetic excitations, or other collective modes. A similar 
formalism can be adopted for the direct contributions to $\ka(T)$ of 
well-defined magnon modes. This type of model has been used to obtain a 
quantitative account of the thermal conductivity in a number of low-dimensional 
spin systems where both the spin excitations and their phonon-scattering 
effects, appearing in the $\tau_{\res}^{-1}$ term, can be characterized 
accurately \cite{Sologubenko2000,Sologubenko2001,Sales2002,Hofmann2001}.

Despite the experimentally verified success of the DC model for 
certain cases, in practical applications the model of Eq.~(\ref{eq:tau}) 
requires at minimum seven to ten fitting parameters to reproduce even rather 
smoothly varying $\ka(H,T)$ curves. The microscopic implications of each term 
are often very difficult to verify independently, unless the respective fitting 
parameters can be compared among closely related materials (for example by 
doping or elemental substitution). Most importantly for the \crcl~problem, 
there are no well-defined spin excitations over most of the ($H,T$) phase 
diagram, and hence no possibility of describing the giant spin-phonon 
scattering within a $\tau_{\res}^{-1}$ term. Clearly the form of this term 
in Eq.~(\ref{eq:tau}) is too restrictive to capture the rich spectrum of 
possible interactions between the phonons and magnetic excitations of a 
quantum magnet, particularly if the latter are fractionalized. Thus we will 
introduce a more general approach to modelling the thermal conductivity of a 
magnetic insulator, a task in which we will be phenomenological but 
quantitative.

The focus of our contribution is to describe the heat conduction of magnetic 
insulators in the common situation where this is governed mostly by phonons, 
but subject to a spin-phonon scattering to which multiple mechanisms may 
contribute. In such a case, it would be highly desirable to have a method 
of understanding the magnetic scattering of phonons without invoking either 
the microscopic details of the scattering mechanism or system-specific 
characteristics such as the phonon and magnon (or spinon) dispersion 
relations. Here we present a phenomenological model to quantify the $T$- and 
$H$-dependence of $\kxx$, for the worked example of \crcl, by considering the 
phonon heat conduction in the presence of scattering by magnetic degrees of 
freedom. The parameters determined empirically in our model enable us to 
quantify the dominant scattering mechanisms regardless of the energy scales 
of the phonons and of the magnetic excitations. 

The structure of this article is as follows. In Sec.~\ref{smm} we summarize 
briefly our samples and experimental methods. In Sec.~\ref{sexp} we show the 
results of our $\ka(H,T)$ observations at all measured fields and compare 
these with measurements of the magnetization and specific heat. In 
Sec.~\ref{smodel} we present the details of our empirical modelling 
procedures for the physically different regimes and extract the quantities 
required to describe spin-phonon scattering in \crcl. Section \ref{sdc} 
provides a discussion and conclusion. 

\section{Material and Methods}
\label{smm}

Thin, purple, plate-like single crystals of \crcl~were grown by chemical vapor 
tranport \cite{Glamazda2017,McGuire2017}. \crcl~is known \cite{Morosin1964} to 
have a rhombohedral low-temperature crystal structure composed of hexagonal 
lattices of Cr$^{3+}$ ions in the $ab$ plane, whose ABC ${\hat c}$-axis 
stacking is ensured by only rather weak (van der Waals) structural 
interactions. Magnetically, as noted in Sec.~\ref{sintro}, the $S = 3/2$ 
(high-spin) Cr$^{3+}$ ions in the honeycomb layers have FM interactions, of 
strength $J = - 5.25$ K \cite{Narath1965}, which is a consequence of the 
near-90$^\circ$ Cr-Cl-Cr geometry (edge-sharing CrCl$_6$ octahedra), while 
the inter-layer interactions ($J'$) are AF.

The magnetization, $m(T)$, and heat capacity, $c(T)$, were measured over a 
range of temperatures from 2 to 100 K and of applied magnetic fields up to 18 
T using, respectively, Quantum Design MPMS and PPMS systems. The in-plane 
longitudinal thermal conductivity, $\kappa_{xx} \equiv \kappa$, was measured 
on an as-grown sample of dimensions 2$\times$4$\times$0.5 mm using a 
single-heater, two-thermometer configuration in steady-state operation with 
the field applied in the $ab$ plane and in the direction of the thermal 
gradient $(\nabla T \parallel \mathbf{H} \in ab)$; limited $\ka$ measurements 
were also performed with the field normal to the plane. The difference in 
absolute temperatures across the sample was set never to exceed 5\% of the 
bath temperature throughout the entire $T$ range of the measurements. All 
thermometry was performed using Cernox resistors, which were pre-calibrated 
individually and {\it in situ} under the maximum applied fields of each 
instrument.

\section{Experimental Measurements}
\label{sexp}

\subsection{Thermal conductivity, $\ka(T)$}

Figure \ref{fig:kT} shows the complete picture of $\ka$ as a function of 
temperature for all values of the applied in-plane field, $H$, that we 
measured. At zero field (ZF), $\ka(T)$ exhibits a somewhat flat maximum 
at 20--25 K with a gentle decline to higher temperatures; at lower $T$ it 
has a marked plateau-type region at and below the magnetic ordering 
temperature, $T_N = 14.2$ K. As $H$ is increased, it is clear that fields 
on the order of $\mu_0H = 1$ T have only a minor effect on $\ka(T)$. However, 
beyond 1 T the applied field causes a dramatic increase of $\ka(T)$ at all 
temperatures and the development of a strong and sharp peak at $T_p \simeq 
20$ K; at 18 T the maximum exceeds its ZF value by a factor of 2.2. As we 
will show below, in fact $\ka(T)$ is almost completely independent of the 
direction of the applied field, indictating a minimal magnetic anisotropy. 

At high fields, the peak at $T_p$ and the line shape on both sides of it, 
which retains only a minor remnant of the plateau at $T_N$, are characteristic 
of phonon-dominated thermal conductivity. The value of $T_p$ varies little as 
$H$ is increased. Its physical origin lies in a crossover between the dominant 
phonon scattering mechanisms. At $T > T_p$, Umklapp processes dominate and 
$\tau_U^{-1}$ is the largest term in Eq.~(\ref{eq:tau}); for $T < T_p$, defect- 
($\tau_{pd})$ and boundary-scattering processes ($\tau_b$) take over. It is 
reasonable to assume that $\ka(T)$ at $\field = 18$ T is closest to 
reproducing the purely phononic response, $\kph(T)$, of \crcl, and we 
return to this topic in detail in Sec.~\ref{smodel}C. As the inset of 
Fig.~\ref{fig:kT} makes clear, at this field the low-$T$ $\kxx(T)$ exhibits 
a $T^3$ dependence, suggesting that the thermal conduction is due to ballistic 
transport of acoustic (linearly dispersive) phonons.

At all lower fields, including zero, $\ka(T)$ reflects a systematic suppression 
due to additional spin-dependent phonon-scattering processes over essentially 
the entire $T$ range. Only at very low temperatures ($T < 5$ K) does the ZF 
$\ka$ exceed that at all other fields, because this is where the contributions 
of coherent magnon excitations in the magnetically ordered phase become 
important. From the inset of Fig.~\ref{fig:kT}, the thermal conductivity in 
this regime has no simply characterized form, and may be a consequence of 
comparable magnon and phonon contributions, at least one of which does not 
show ballistic transport \cite{Pohl1982}. By contrast, at all temperatures 
above 5 K the dominant effect of the spins is not an additive contribution 
from three-dimensionally coherent excitations but a destructive effect due 
to scattering of the phonons by incoherent spin fluctuations. It is the 
field-induced suppression of these fluctuations that brings about the 
striking enhancement we observe in $\ka$. 

\begin{figure}[t]
\includegraphics[width=\linewidth]{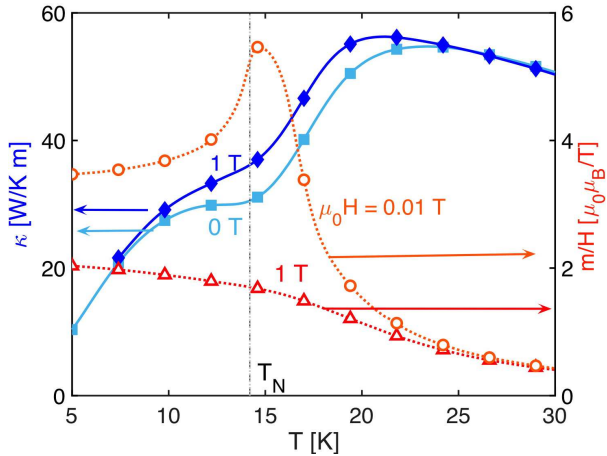}
\caption{Data for $\ka(T)$ at 0 and 1 T reproduced for comparison with 
the field-normalized magnetization, $\chi(T) = m/H$. The very strong peak 
in the low-field $\chi(T)$, indicative of strong fluctuations towards 
magnetic order around $T = T_N$, matches the plateau in $\ka(T)$ where its 
growth is arrested.}
\label{fig:kTlowfield}
\end{figure}

\subsection{Magnetization}

To understand this interplay of lattice and magnetic excitations, we first 
consider the magnetic properties of \crcl. The magnetization, $m(H)$, is 
shown in the inset of Fig.~\ref{fig:nH} for several different temperatures. 
As noted in Sec.~\ref{sintro}, a moderate field $\mu_0 H \approx$ 2 T, applied 
in the plane, is sufficient at $T < 5$ K to rotate the FM planes against the 
AF interplane interaction and drive the system to saturation. We comment that 
the field scale for this process suggests, in contradiction to the conclusion 
of Ref.~\cite{Narath1965}, that the interplane $J'$ is of order 1 K. We deduce 
a saturation moment, $m_s = m(T \rightarrow 0)$, of 2.88$\mu_B$ per Cr$^{3+}$ 
ion. The same value of $m_s$ is obtained by applying the same field in the 
out-of-plane direction, demonstrating that in \crcl, unlike \rucl, the 
magnetic anisotropy is extremely weak \cite{McGuire2017}. This $m_s$ value 
is fully consistent with the spin contribution expected for a single Cr$^{3+}$ 
ion of $S = 3/2$ when the $g$-factor appropriate for a magnetically isotropic 
system, $g = 2$, is assumed. 

For perspective on the relation between $\ka(T)$ and $m(T)$, in 
Fig.~\ref{fig:kTlowfield} we show $\ka(T)$ on the same temperature axis as 
the field-normalized magnetization, $m(T)/H \equiv \chi(T)$, for fields of 0 
and 1 T in $\ka$ and 0.01 and 1 T in $\chi$. This low-field regime is where 
the plateau in $\ka(T)$ around $T_N$ is most pronounced, and the field is not 
sufficient to suppress the spin-induced scattering of the phonons conducting 
the heat. For very small fields, it is clear that $\chi(T)$ becomes very 
large in the region of the ordering transition, indicating strong magnetic 
fluctuations and hence the origin of the $\ka$ plateau. This behavior is 
suppressed considerably at 1 T, where the plateau begins to lose its form. 
Above 25 K, the residual $m$ scales precisely with $H$. 

\begin{figure}[t]
\includegraphics[width=\linewidth]{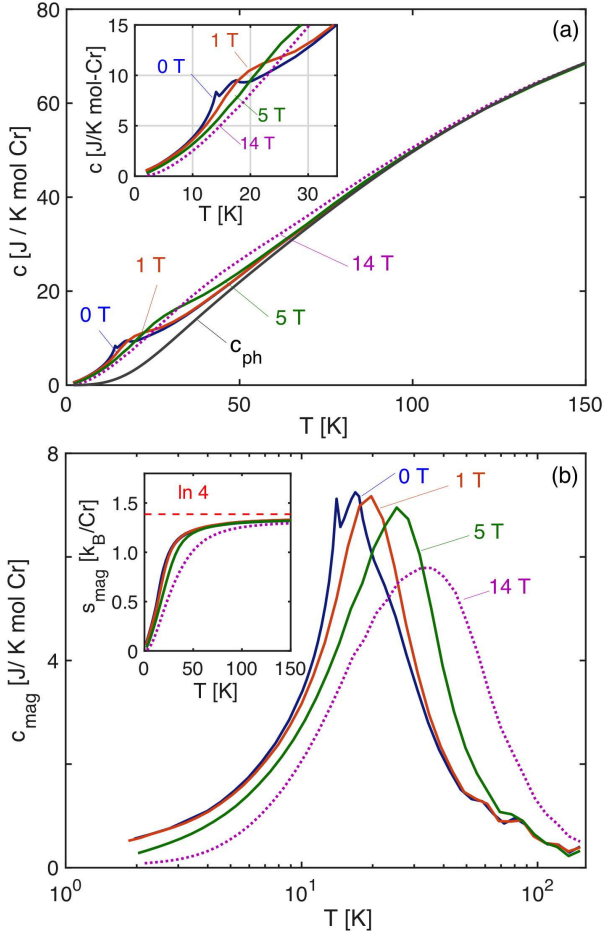}
\caption{(a) Specific heat, $c(T)$, for a selection of different applied 
magnetic fields. The black line shows our estimate of the phonon contribution, 
$\cph(T)$, obtained from a Debye-model fit to $c(T)$ data for the non-magnetic 
analog ScCl$_3$ \cite{HwanDo2017}. (b) Estimated magnetic specific heat,  
$\cmag(T)$, obtained by subtracting $\cph(T)$ from $c(T)$ at each field. The 
inset shows the magnetic entropy, $s_{\rm mag}(T)$, which in the high-temperature 
limit approaches the value $k_B \ln 4$.}
\label{fig:Cdept}
\end{figure}

\subsection{Magnetic specific heat}

To obtain further insight into both the phonons and their spin-mediated 
scattering, next we consider the specific heat as a function of $T$ and 
$H$. Figure \ref{fig:Cdept}(a) shows our measurements of the total specific 
heat, $c(T)$, which are fully consistent for all $H$ with previously reported 
results \cite{McGuire2017}. To estimate the phonon contribution, $\cph(T)$, 
we employ a three-dimensional (3D) Debye-model fit of heat-capacity data 
for the non-magnetic analog \sccl~\cite{HwanDo2017}. The resulting 
$\cph(T)$, shown by the green line in Fig.~\ref{fig:Cdept}(a), agrees well 
with our data for \crcl~in the high-temperature regime, which as the inset 
illustrates is $T \gtrsim 100$ K.  

We attribute the remaining, strongly field-dependent portion of the specific 
heat to the magnetic degrees of freedom, meaning we define $\cmag(T) = c(T)
 - \cph(T)$. Figure \ref{fig:Cdept}(b) shows $\cmag(T)$ on a logarithmic 
temperature axis. At ZF there are clearly two peaks, a sharp, $\lambda$-type 
anomaly at the N\'eel transition, $T_N = 14.2$ K, and a rounder maximum at 
17-18 K. When the applied magnetic field is increased to 1 T, the $\lambda$ 
anomaly is completely lost, and in fact a detailed study of the low-field 
evolution of this feature \cite{McGuire2017} found that a field of 0.2 T is 
sufficient to suppress this indicator of the ordering transition. As the 
applied field is increased, the location, $T_{\rm max}(H)$, of the broad maximum 
shows an abrupt initial shift towards higher temperatures, before increasing 
more slowly with $\mu_0 H$ to a value of 36 K at 14 T. 

Quite generally, the broad maximum in $\cmag(T)$ fingerprints the energy 
scale of the dominant local spin-flipping processes in the system. In \crcl, 
this is the energy cost for reversing a single spin in the FM honeycomb layers 
(in the rhombohedral structure one would anticipate an energy of $3 |J| + J'$). 
The initial increase in $T_{\rm max}(H)$ can be understood as a straightforward 
reinforcement of this magnetic stiffness while the applied field competes 
with $J'$ to reorient the FM layers. The slower increase at $\mu_0 H > 2$ T 
corresponds to a competition of the field with $3|J|$, which in 
Ref.~\cite{McGuire2017} was formulated as a progressive development of 
ferromagnetic correlations in the plane. Because of the low spin-coercivity 
in an applied field, FM alignment of very large domains becomes 
thermodynamically favorable at higher $T$. It is important to note that 
$T_N$, the temperature for 3D magnetic order, almost coincides with the 
broad maximum at ZF. Thus despite the quasi-2D nature of the structure 
of \crcl, the AF state at ZF is magnetically 3D, with $T_N$ of the same 
order as the Curie-Weiss temperature ($\theta_{\rm CW} \simeq 30$ K both from 
our data below the structural transition at 240 K and from that of 
Ref.~\cite{McGuire2017}). 

\begin{figure*}[t]
\includegraphics[width=\linewidth]{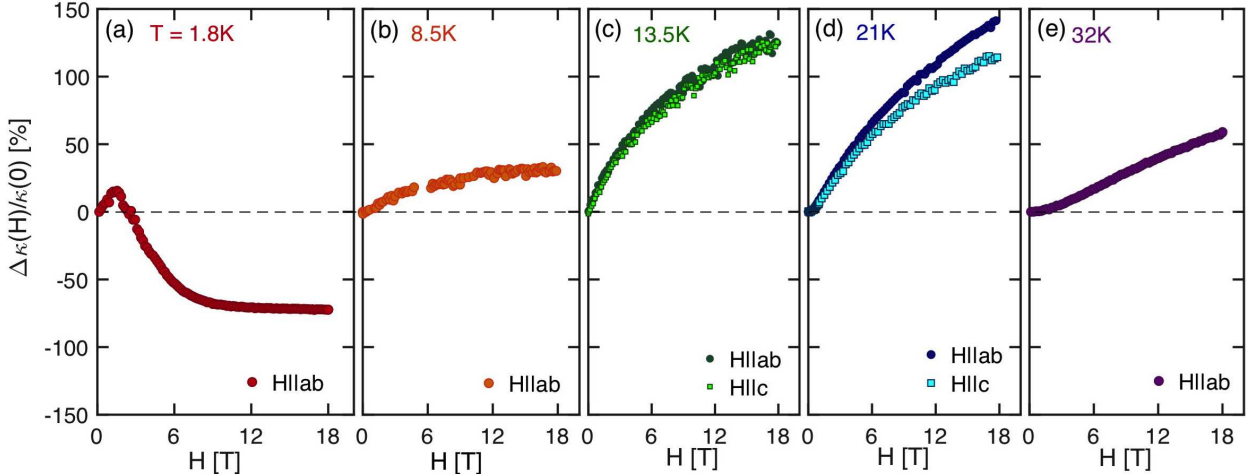}
\caption{Fractional change in $\kxx(H)$. For each selected temperature value 
we define $\delk/\kappa_0 = [\kxx(H) - \kxx(0)]/\kxx(0)$. $H$ is applied 
in the $ab$ plane in every panel, and also along the $c$-axis at $T =$ 13.5 
and 21 K, which highlights the very low magnetic anisotropy. For all 
temperatures $T > 4$ K, the applied field causes a monotonic enhancement of 
$\ka$. Only at $T < 4$ K is a different behavior observed as a consequence of 
the coherent magnonic contribution: $\kxx$ is enhanced by a small applied 
field (which causes magnon stiffening) but suppressed by all higher fields 
(where the energy cost of magnon excitation becomes prohibitive).} 
\label{fig:kH}
\end{figure*}

While one may fear that our estimated $\cmag(T)$, as a small difference 
between two larger numbers, is subject to significant errors, a complete
validation of its accuracy may be obtained by integration. The magnetic 
entropy, $s_{\rm mag}(T) = \int_0^T dT' \, c(T')/T'$, is shown in the inset of  
Fig.~\ref{fig:Cdept}(b). Clearly at temperatures beyond 50 K, $s_{\rm mag}(T)$ 
for all fields displays a smooth and accurate approach to the limit $k_B \ln 
4$ expected for the four-level system corresponding to free $S = 3/2$ spins. 
This result, which contrasts strongly with the entropies shown in 
Ref.~\cite{McGuire2017}, indicates that our approach of deducing the 
lattice contribution from the material ScCl$_3$ provides a quantitatively 
accurate estimate of $\cph(T)$.  

\subsection{Thermal conductivity, $\ka(H)$}

The clearest way to gauge the effects of the magnetic state on the thermal 
conductivity is to consider the isothermal $H$-dependence of $\kxx$. In  
Fig.~\ref{fig:kH} we show for a range of temperatures the fractional change 
$\delk/\kappa_0 = [\kxx(H) - \kxx(0)]/\kxx(0)$ caused by an applied magnetic 
field. It is apparent immediately that the generic field response is a 
monotonic increase in $\delk/\kappa_0$, except at the lowest temperatures. 

Although our $T = 1.6$ K data are something of an outlier for our present 
purpose, which is to discuss the generic behavior beyond the ordered regime, 
we focus first on the low-$T$ case. Here $\kxx(H)$ increases initially, peaking 
around 2 T before decreasing to negative values of $\delk/\kappa_0$ and becoming
$H$-independent at fields $\field > 8 $ T. This behavior, which is observed 
for $T < 4$ K, indicates a coherent magnonic contribution to heat conduction 
that is comparable to the phononic one. The coherent spin-sector contribution 
to thermal conductivity has been the focus of numerous studies \cite{Ando1998,
Sologubenko2000,Sales2002,Jin2003,SYLi2005,Leahy2017} and can in general be 
described as an independent additive term beyond the phonon contribution, 
i.e.~$\kappa_{\rm tot} = \kph + \kmag$. At low fields, the initial rise of 
$\kxx(H)$ can be explained by the magnon stiffening that occurs when 
$\field$ overcomes $J'$ to orient all of the layers ferromagnetically, 
which results in an increase of the magnon propagation speed and hence 
of $\kmag$ \cite{Hirschberger2015}. However, this contribution decreases as 
soon as the field-induced magnon gap exceeds the measurement temperature, at 
which point the magnon population is suppressed exponentially, leaving a 
smaller and largely $H$-independent thermal conductivity, $\kappa_{\rm tot} 
\rightarrow \kph$. The coherent magnon contribution also diminishes quickly 
as $T$ increases, and in fact $\delk/\kappa_0$ changes character well below 
$T_N$ (e.g.~$T = 8.5$ K in Fig.~\ref{fig:kH}). 

At all temperatures $T > 4$ K, magnetic fluctuations play the role of 
scattering centers, rather than participants in coherent heat conduction. 
While this type of behavior has been observed in certain quantum magnetic 
materials at specific temperatures, where it can be described by a $\tau_{\res}
^{-1}$ term in Eq.~(\ref{eq:tau}) \cite{Hofmann2001,Sales2002,Wu2015}, 
systems in which it appears across the full range of temperatures are not 
widely known. Nonetheless, it is clear in all the relevant panels of 
Fig.~\ref{fig:kH} that $\ka$ increases monotonically with field; we remark 
again that this behavior is largely independent of the field direction. While 
it remains the case that increasing $H$ suppresses the spin fluctuations, its 
effect is a uniform suppression of the spin-phonon scattering by an effective 
reduction in the density of scattering centers, thus bringing the system 
closer to purely phononic heat conduction.

\section{Empirical Model for $\kappa(H,T)$}
\label{smodel}

To model the thermal conductivity in the presence of such a large and 
obviously destructive spin-phonon interference effect, we base our analysis 
on the thermal conductivity, $\kph(T)$, due to phonons. We first reexpress 
Eq.~(\ref{eq:tau}) in the form 
\bee
\tau^{-1} = \tau_0^{-1} + \tau_{\rm mag}^{-1},
\label{etau}
\ene 
i.e.~we assume that the effective phonon scattering rate can be separated 
into a part $\tau_0^{-1} \equiv \tau_0^{-1}(T)$ containing all the 
field-independent terms in Eq.~(\ref{eq:tau}) and a part $\tau_{\rm mag}^{-1} 
\equiv \tau_{\rm mag}^{-1}(H,T)$ containing all phonon scattering processes 
involving magnetic degrees of freedom. 

To capture the behavior of $\kxx(H,T)$ phenomenologically, we assume 
that the relative scattering rate, $\tau_{\rm mag}^{-1}/\tau_0^{-1}$, will depend 
on the population of magnetic fluctuations, which we denote $\nmag$, and on 
a dimensionless coupling constant describing the strength, or effectiveness, 
of the phonon scattering processes, $\lambda(H,T)$. Thus the key equation 
underpinning our empirical treatment is that the thermal resistivity will be 
given by 
\bee
\wxx \! (H,T) = \wph(T) [1 + \lambda(H,T) \, \nmag(H,T)],
\label{eq:wxx}
\ene
where $\kph$, the phonon thermal conductivity in the absence of spin-phonon 
interactions, is $H$-independent and would be recovered at high applied 
fields. By some straightforward algebra, one may verify that $\lambda \nmag 
 = \tau_{\rm mag}^{-1}/\tau_0^{-1} = \tau_0/\tau_{\rm mag}$. We normalize $\nmag$ 
to the total number of Cr$^{3+}$ spins so that it represents a fractional 
density, and thus the field-dependence of $\kxx$ is encoded in two unitless 
parameters, $0 < \nmag < 1$ and $0 < \lambda$.

\begin{figure}[t]
\includegraphics[width=\linewidth]{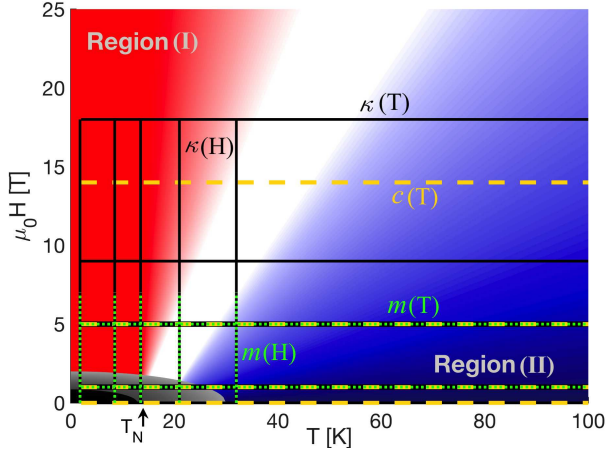}
\caption{Schematic representation of the $(H,T)$ phase diagram of CrCl$_3$.
Graded red colors denote the extent to which the system is in the regime of 
high $H/T$ [region (\textbf{I})] and blue colors the regime of low $H/T$ 
[region (\textbf{II})]. The black color denotes the regime of long-ranged 
3D AF order and the gray colors the extent to which the approximations we 
apply in regions (\textbf{I}) and (\textbf{II}) are invalidated by the 
residual presence of short-ranged magnetic correlations. Solid black lines 
denote the fields and temperatures over which we have measured the $\ka(T)$ 
data shown in Fig.~\ref{fig:kT} and the $\ka(H)$ data shown in 
Fig.~\ref{fig:kH}, dashed orange lines the specific-heat data shown in 
Fig.~\ref{fig:Cdept}, and dotted green lines the $m(H)$ and $m(T)$ data 
shown respectively in Figs.~\ref{fig:nH} and \ref{fig:AFM_MvT}.}  
\label{fig:5n}
\end{figure}

Focusing now on the spin fluctuations responsible for phonon scattering, 
because these change character across the ($H,T$) phase diagram, we begin by 
subdividing this into the two regimes shown schematically in Fig.~\ref{fig:5n}. 
Qualitatively, at high fields and low temperatures one expects the fluctuations 
to be well-defined but 2D spin-wave excitations within the FM layers, whereas 
for temperatures high relative to the field one expects random fluctuations of 
weakly coupled spins. Specifically, for \crcl~we define region (\textbf{I}) as 
covering low temperatures ($T < T_N$) for fields $H \ge 2$ T and $T < 30$ K at 
our highest measurement field; here we will find that $n_{\rm mag}$ corresponds 
closely to the density of magnon excitations originating within the honeycomb 
layers, which are highly spin-polarized and strongly ferromagnetically 
correlated, and hence remain coherent 2D entities outside the AF phase. 
Region (\textbf{II}) covers temperatures $T > T_N$ for low fields, and at 
our higher fields temperatures $T > 40$ K; here $n_{\rm mag}$ corresponds to the 
average density of free spins that are not aligned with the field direction, 
and as such is well described by a Weiss-field picture, within which the net 
magnetization is determined by conventional paramagnetic behavior. As 
Fig.~\ref{fig:5n} makes clear, the AF ordered state occupies a very small 
region of the $(H,T)$ phase diagram, and because the contributions of coherent 
3D magnon excitations to $\ka$ are also small (Fig.~\ref{fig:kT}), the ordered 
regime is not the focus of our study. However, we comment that short-ranged 
magnetic correlations in the vicinity of the ordered state (gray colors) will 
act to limit the applicability of the approximations we apply below. 

By considering the magnetic scattering in the high-$H/T$ (\textbf{I}) and 
low-$H/T$ regions (\textbf{II}) of Fig.~\ref{fig:5n}, we will minimize the 
ambiguity at intermediate $H/T$ ratios. In both regions, the fractional 
density of magnetic fluctuations, either of 2D spin-wave excitations or of 
freely fluctuating spins, can be fixed accurately to known limits. In region 
(\textbf{I}) this limit is the fractional deviation, $1 - m(H,T)/m_s$, of the 
net magnetization from its high-$H/T$ saturation value, which is $m_s \equiv 
m(H \rightarrow \infty)$. In region (\textbf{II}) it is the fractional 
polarization, $m(H,T)/m_s$, which vanishes in the low-$H/T$ limit. 

It is apparent from Eq.~(\ref{eq:wxx}) that the type of phenomenological 
approach we adopt requires a reliable estimate of the pure (field-independent) 
phonon thermal conductivity, $\kph(T)$, to be useful or even viable. If the 
dependence of $\ka$ on the applied field is sufficiently weak at strong 
fields, many authors \cite{Jin2003,Pan2013} use the $\kxx(T)$ data at their 
highest available magnetic field as a measure of $\kph$. In Sec.~\ref{smodel}C 
we will use our modelling procedure to obtain a field-independent $\kph(T)$ by 
extrapolating from our $\ka(H,T)$ measurements (shown in Fig.~\ref{fig:kT}), 
and hence will determine how close our $\ka (\mu_0H = 18$ T) data are to 
saturation. 

\begin{figure}[t]
\includegraphics[width=\linewidth]{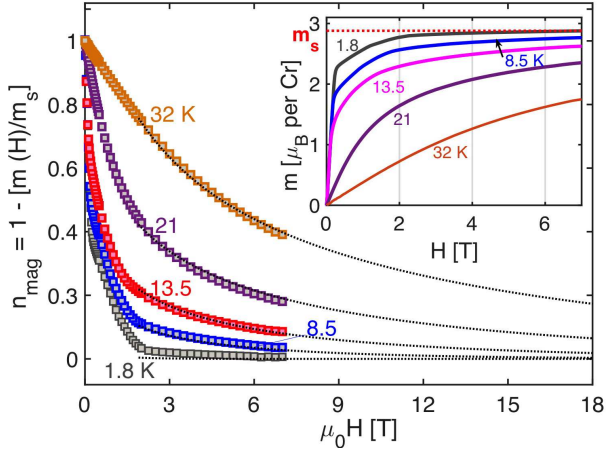}
\caption{Estimate of the population of magnetic scattering centers, $n_{\rm mag}
(H) = 1 - m(H)/m_s$, taken from the fractional deviation of the magnetization, 
$m(H)$, measured in the MPMS at several fixed low temperatures from its 
high-$H$ saturation value. The black dotted lines show the same quantity 
computed for 2D magnetic excitations on a FM honeycomb lattice, using 
Eq.~(\protect{\ref{eqn:LSW}}) with $d = 2$; this approach can be extended far 
beyond the range of the available data. Inset: isothermal magnetization, 
$m(H)$, for fields up to 7 T, shown for the same temperatures.}  
\label{fig:nH}
\end{figure}

\subsection{Region (\textbf{I}): dominant magnon scattering}

Our magnetization data (inset Fig.~\ref{fig:nH}) show that, for temperatures 
$T < 30$ K, fields $\mu_0H \ge 5$ T are strong enough to polarize more than 
50\% of the Cr$^{3+}$ spins. This temperature and field range lie well within 
the ``spin-flop'' regime \cite{Narath1965,McGuire2017}, where the spins are 
forced into a quasi-FM-ordered state of the honeycomb layers. To model this 
regime, we assume 2D FM spin-wave excitations with the field-gapped 
dispersion relation $\hbar \omega (\mathbf{k}, H) = {\textstyle \frac12} 
{\tilde J} (k a)^2 + g\mu_B H$, where $a$ is the lattice constant and 
${\tilde J}$ is an effective magnetic interaction strength. The population 
of magnetic excitations per unit volume in $d$ dimensions may be then 
estimated using the expression
\begin{align}\label{eqn:LSW}
n_{\rm mag} (H,T) & = \int \dfrac{d^dk}{(2\pi)^d} \dfrac{1}{e^{\hbar\omega 
(\mathbf{k},H)/k_BT} - 1}, \\ & = \alpha \left( \dfrac{k_BT}{\tilde J} 
\right)^{d/2} {\rm Li}_{d/2} \left( e^{-g\mu H/k_BT} \right), \nonumber  
\end{align}
where we assume that only wave vectors $\mathbf{k}$ near the zone center 
contribute significantly to $n_{\rm mag}$, and thus we take the upper limit of 
the integral to infinity. $\alpha$ is a unitless constant set by the dimension, 
$d$, and the normalization. Li$_{s}(x)$ is the polylogarthmic function of order 
$s$, which has an infinite series expression that is easy to evaluate 
numerically. As Fig.~\ref{fig:nH} makes clear, this approach provides an 
excellent account of the $H$- and $T$-dependence of $m$, effective even at 
temperatures (21 and 32 K) outside region (\textbf{I}) as it is represented 
in Fig.~\ref{fig:5n}. In Fig.~\ref{fig:nH} the only fitted parameter is the 
effective interaction, ${\tilde J} = 13.1$ K; clearly ${\tilde J} \approx 3 J$ 
is again close to the characteristic energy scale of the 2D FM layers at ZF. 
Here we have taken ${\tilde J}$ to be constant, because one does not 
anticipate that fields up to 7 T could compete significantly with the 
in-plane energy scale (this is to be contrasted with the ``magnon stiffening'' 
discussion in Secs.~\ref{sexp}C and \ref{sexp}D, where $\field$ competes with 
$J'$). We stress again that, as shown in Fig.~\ref{fig:kT}, these magnonic 
fluctuations are a significant source not of heat conduction but of phonon 
scattering.

\begin{figure}[t]
\includegraphics[width=\linewidth]{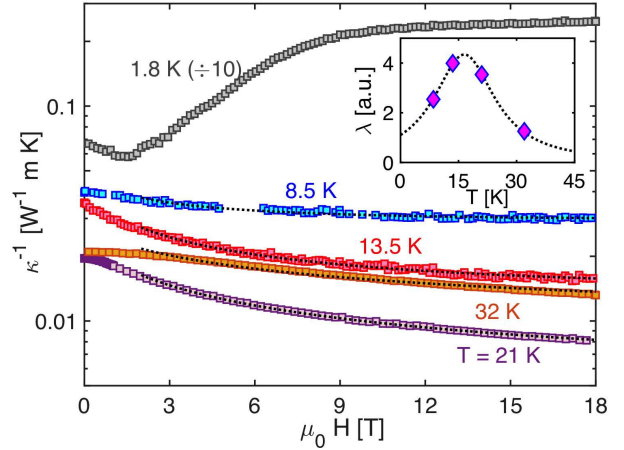}
\caption{Isothermal thermal resistivity, $\kappa^{-1}(H)$, obtained from 
Fig.~\ref{fig:kH}. The dotted black lines are fits of the data at $\mu_0H > 2$ 
T to the form of Eq.~(\ref{eq:wxx}), from which we deduce values for the 
high-field limit [the phonon thermal resistivity, $\kph^{-1}(T)$], and the 
dimensionless and field-independent coupling strength, $\lambda (T)$. The 
inset shows the values of $\lambda(T)$ deduced at each temperature.}
\label{fig:wH}
\end{figure}

At high $H/T$, when the total magnetization remains a significant fraction 
of $m_s$, we equate the measured deviation directly with $n_{\rm mag}$, the 
population of 2D magnons obtained using Eq.~(\ref{eqn:LSW}) with $d = 2$. 
The agreement remains good for all temperatures $T < 2T_N$ when $\mu_0 
H \ge 2$ T. Given the discrepancy between the in-plane FM interactions and 
the out-of-plane AF ones, whose 3D coupling effects are suppressed beyond 2 T 
(inset Fig.~\ref{fig:nH}), it is not surprising that quasi-2D excitations 
dominate the spectrum. We comment here that the CrX$_3$ materials have been 
considered as candidates for hosting topological magnons with Dirac cones 
in their (graphene-type) dispersion relations \cite{Pershoguba2018}, but at 
finite applied fields a full gap is opened and the magnon dispersion has the 
quadratic form assumed in our model. 

To analyze the phonon scattering strength, $\lambda(H,T)$, in Fig.~\ref{fig:wH} 
we use Eq.~(\ref{eq:wxx}) to fit the thermal resistivity data, $\wxx(H)$, for 
four of the five temperatures shown in Fig.~(\ref{fig:kH}). We find that an 
excellent description is obtained with a function $\lambda(H,T) \equiv 
\lambda(T)$ completely independent of the field. Thus in region (\textbf{I}) 
the field-dependence of $\ka(H,T)$ is contained entirely within our 
straightforward assumptions about $\nmag(H,T)$. Further, we observe that 
$\lambda(T)$ peaks in temperature close to $T_N$, and thus we deduce that 
scattering by coherent but ``only'' 2D magnetic excitations has its strongest 
effect on the phonon thermal conductivity in the regime around $T_N$ itself; 
from the form of the magnetic specific heat [Fig.~\ref{fig:Cdept}(b)], it is 
no surprise that phonon scattering should be most efficient around this 
characteristic temperature. This leads to a suppression effect that includes 
the peak region (around $T_p \approx 20$ K) and becomes unusually large as 
the applied field is reduced below 2 T, where $n_{\rm mag}$ rises strongly 
(Fig.~\ref{fig:nH}). We note that, even in a van der Waals material, the 
phonons are in general much more 3D in nature than are the magnons once their 
interlayer correlations have been destroyed by the applied field, as a result 
of which good 2D magnons with no interlayer coherence are only damaging to 
phonons, and hence to thermal conduction.

\subsection{Region (\textbf{II}): dominant paramagnetic fluctuations}

At higher temperatures, the magnetic fluctuations may no longer be regarded 
as well-defined 2D spin-wave excitations. As the energy scale of thermal 
fluctuations becomes large compared to that of the field, planar FM order is 
destroyed and the behavior of the spin system becomes paramagnetic. Despite 
this loss of coherent 2D magnons, magnetic scattering continues to play a 
significant role in controlling the thermal conductivity in region 
(\textbf{II}), as Fig.~\ref{fig:kT} shows clearly. This effect must be a 
consequence of phonon scattering off randomly fluctuating free paramagnetic 
spins, whose relative density is also given by the fractional deviation of 
the magnetization from $m_s$.

\begin{figure}[t]
\includegraphics[width=\linewidth]{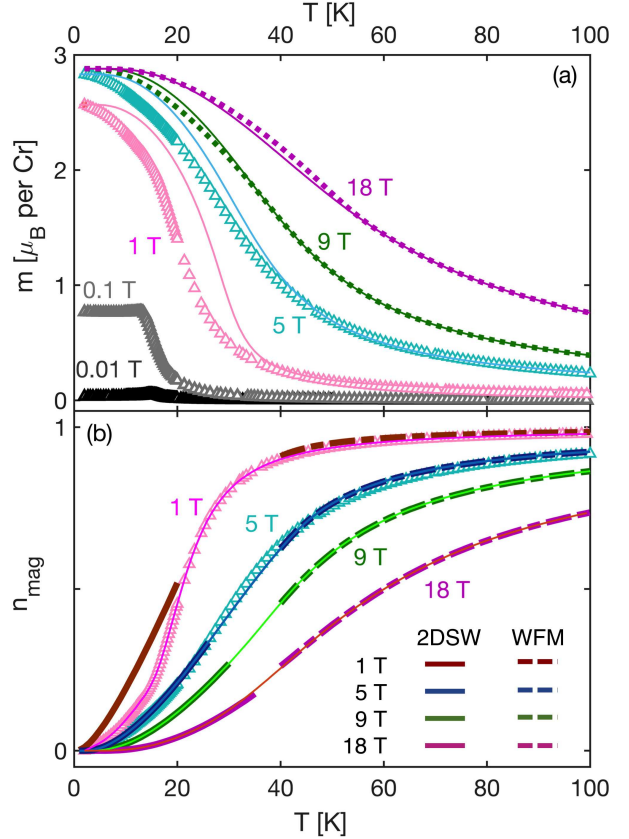}
\caption{(a) Magnetization, $m(T)$, measured at fixed field values of 
0.01, 0.1, 1, and 5 T. Thin solid lines are fits made by applying the 
Weiss-field model [Eq.~(\ref{eq:MT})] at $T > 40$ K. This model can be 
extrapolated to $T = 0$, although of course it does not include the AF magnetic 
order, and this accounts for the deviations at $T < 40$ K for low fields. In 
addition we show calculations made with the same model parameters for $m(T)$ 
at 9 and 18 T, beyond the range of the data, where deviations due to AF order 
would be very small. (b) $\nmag(T)$ deduced from $m(T)$; here we show the 
quantity $\nmag (H,T) = 1 - m(H,T)/m_s (H,T = 0)$ at each constant value of 
$H$. Solid lines marked 2DSW are estimates of $\nmag(T)$ obtained from the 2D 
spin-wave model of Sec.~\ref{smodel}A. Dashed lines marked WFM are estimates 
of $\nmag(T)$ obtained from the Weiss-field model of Sec.~\ref{smodel}B. Solid 
lines for 5, 9, and 18 T mark spline fits connecting the two regimes; at 1 T 
the 2DSW approach is not effective because it does not contain the magnetic 
order, and here the solid line is a fit to the data.}
\label{fig:AFM_MvT}
\end{figure}

Although the dominant physics in region (\textbf{II}) is thermal fluctuations 
that flip individual spins against the magnetic field, the underlying spin 
interactions may by no means be neglected. We model the magnetization profile, 
$m(T)$, of \crcl~by a Weiss-field approach \cite{Blundellbook} in which the  
molecular-field term takes into account the FM correlations within the 
honeycomb planes. In this framework 
\bee
m(H,T) = m_s B_S \! \left[ \dfrac{g \mu_{\rm B} S}{k_BT} \!\!\left( \mu_0 H + 
B_{\rm mol} \dfrac{m(T)}{m_s} \right) \! \right] \!, 
\label{eq:MT}
\ene
where $B_S(y)$ is a Brillouin function of order $S$, $S = 3/2$, and $g = 2$.   
The solution for $m(T)$ is found by fixed-point iteration and, as shown in 
Fig.~\ref{fig:AFM_MvT}(a), an exceptionally good fit to the measured 
magnetization is obtained over the entire paramagnetic temperature range, 
$T \ge 40$ K, for all fields below 7 T. The fitted constant $B_{\rm mol} = 22.4$ 
T $\simeq$ 15 K is given once again by the FM in-plane coupling scale. In 
addition to providing smooth and easily computed curves at all measured 
field strengths, this model allows us to predict $m(T)$ with high confidence 
over the same temperature range at fields of $\mu_0H = 9$ and 18 T that are 
prohibitively high for SQUID magnetometer measurements (Fig.~\ref{fig:AFM_MvT}).
The clear deviations between model and measurement at low fields and 
temperatures are consequences of the AF order, and are suppressed by fields 
in excess of 1 T (inset Fig.~\ref{fig:nH}), to the point where deviations at 
the fields we can only model ($\mu_0H = 9$ and 18 T) are expected to be 
negligible. 
 
To estimate the scattering strength, $\lambda(H,T)$ in Eq.~(\ref{eq:wxx}), in 
region (\textbf{II}), we observe first that when $T \to \infty$ the magnetic 
degrees of freedom become entirely disordered, meaning that $\nmag \to 1$. 
Thus to preserve the empirical observation that the field-dependence of 
$\kxx(T)$ disappears at high $T$ (Fig.~\ref{fig:kT}), it is necessary that 
$\lambda(H,T) \to 0$ as $T \to \infty$. On physical grounds, this must be 
the case because different phonon scattering mechanisms will overwhelm the 
magnetic contribution in Eq.~(\ref{etau}), rendering $\tau_{\rm mag}^{-1} \ll 
\tau_0^{-1}$. However, from the Weiss-field estimate of $m(H,T)$ shown in 
Fig.~\ref{fig:AFM_MvT}, it is clear that this regime is reached only at 
our lowest measurement fields, and that $\nmag(H,T)$ remains significantly 
less than unity over the available temperature range for all fields above 
1 T. 

From experiment we have been able to obtain four highly accurate estimates 
of $\lambda(T)$ at temperatures up to $T = 32$ K, as shown in the inset of 
Fig.~\ref{fig:wH}. Because the function $\lambda(T)$ is already well 
past its peak value, we use the logic of the previous paragraph to adopt 
test functions for the form of the vanishing of $\lambda(T)$ in region 
(\textbf{II}), taking for specificity Gaussian, exponential, and power-law 
(Lorentzian) forms. Stated briefly, all forms of this functional tail give 
equally valid fits to the data throughout region (\textbf{II}), and we pursue 
the quantitative aspects of this assertion in conjunction with the extraction 
of the phonon thermal conductivity, $\kph(T)$, in the next subsection. The 
key qualitative point to be made here is that any $H$-dependence of the 
functional tail is so weak that, once again, a completely adequate fit to 
all data is obtained by taking $\lambda(H,T) \equiv \lambda(T)$, exactly 
as in region (\textbf{I}). Thus we have obtained the profound result that, 
to a very good approximation, the effect of a magnetic field on the thermal 
conductivity may be encoded entirely within the number of magnetic 
fluctuations, $n_{\rm mag}(H,T)$, acting to scatter the phonons transporting 
the heat, while their scattering efficiency is effectively $H$-independent. 

\subsection{Phonon Thermal Conductivity}

In the preceding subsections we have shown that is it possible to obtain 
excellent descriptions of $m(T)$, and hence $\nmag(T)$, in both regions 
(\textbf{I}) and (\textbf{II}), as shown respectively in Figs.~\ref{fig:nH} 
and \ref{fig:AFM_MvT}, by using minimal physical assumptions. Even without 
a quantitative matching of our two treatments, it is possible on this basis 
to obtain accurate fits of $\ka^{-1}(H,T)$ at all fields and temperatures 
(Fig.~\ref{fig:wH}) by using Eq.~(\ref{eq:wxx}) with a field-independent 
scattering strength, $\lambda(T)$. The exercise of matching the models of 
Secs.~\ref{smodel}A and \ref{smodel}B across the intermediate temperature 
regime would involve the extrapolation of either into a parameter range 
where it is explicitly no longer valid. However, the 2D spin-wave model 
applied in region (\textbf{I}) does account correctly for the temperature 
regime around the peak in $\ka(T)$ for all fields in our measurement range, 
and thus this matching takes place only on the high side of the peak.

It is clear from Eq.~(\ref{eq:wxx}) that $\ka(T)$ is governed largely by 
the ``continuous parameter'' $\kph(T)$, and as such that the most systematic 
matching of the two regimes would be ensured by obtaining an optimal estimate 
of this quantity. $\kph(T)$, as the thermal conductivity due only to phonon 
contributions, is in principle obtained in region (\textbf{I}) as the 
$H \to \infty$ limit, where the spin excitations have an infinite gap and 
$\nmag \to 0$, whence $\ka(T) \to \kph(T)$. In Sec.~\ref{sintro} we summarized 
the difficulties in obtaining $\ka(T)$ from a DC formalism \cite{Callaway1959,
Bergmanbook}, and in fact these are manifest even in obtaining $\kph(T)$. 
Despite the many parameters, the form of the available relaxation-time 
terms remains highly constraining; in the present case, in the absence of a 
non-resonant spin-phonon suppression term, it is not possible to satisfy the 
low- and high-$T$ limits simultaneously with an accurate estimate of the peak 
position, $T_p$. An approach simpler than the full DC treatment is to proceed 
from the expression \cite{Zimanbook} 
\bee
\!\!\! \kph(T) = \tfrac{1}3 c_{\rm ph}(T) v_s \ell_{\rm eff}(T) = \tfrac{1}3 
c_{\rm ph}(T) v_s^2 \tau_{\rm eff}(T),
\label{ekph}
\ene 
where $c_{\rm ph}(T)$, the heat carried by a phonon mode, is known, $v_s$ is 
again an average phonon velocity, and a simple model can be constructed for 
the effective mean free path, $\ell_{\rm eff}$, whose corresponding effective 
scattering time, $\tau_{\rm eff}(T)$, may be connected to Eq.~(\ref{etau}). 
However, once again it is not possible within such a simplified framework 
to obtain an accurate value of $T_p$, and hence $\ka(H,T)$ cannot be 
reproduced with any quantitative accuracy. 

\begin{figure}[t]
\includegraphics[width=\linewidth]{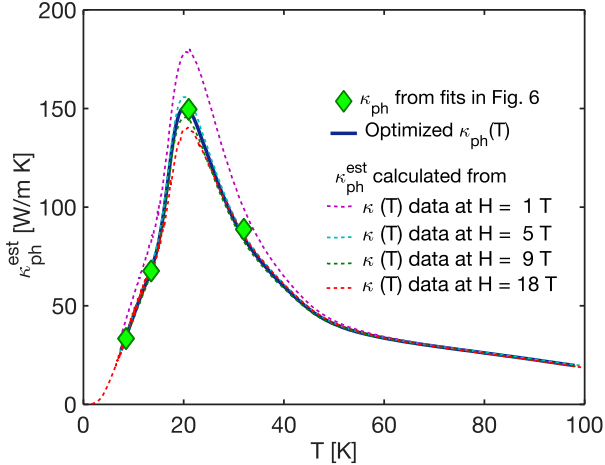}
\caption{Estimates of the field-independent lattice contribution, 
$\kph^{\rm est} (T)$ (dashed lines), obtained by using Eq.~(\ref{eq:wxx}) with 
our measured thermal conductivity data (Fig.~\ref{fig:kT}) at constant fields 
of 1, 5, 9, and 18 T. Green diamonds show values of $\kph$ obtained from fits 
of the isothermal $\ka(H)$, shown in Fig.~\ref{fig:wH}, at temperatures of 
8.5, 13, 21, and 32 K. The thick solid line shows the optimal $\kph(T)$  
deduced from the $\kph^{\rm est}(T)$ curves (see text).}
\label{fig:KvTph}
\end{figure}

To ensure full generality of our treatment and to avoid the pitfalls inherent 
in adopting any approximate forms, we proceed instead directly from our data 
to compute the estimates $\kph^{\rm est} = \ka(H,T) [1 + \lambda(T) \nmag(H,T)]$
for each available field value. Our fits of $\wxx(H)$ in region (\textbf{I}), 
shown in Fig.~\ref{fig:wH}, provide reliable estimates of $\kph$ at four 
discrete temperatures, which are shown as the green diamonds in 
Fig.~\ref{fig:KvTph}. Inverting our $\ka(H,T)$ data requires quantitative 
estimates of $\lambda(T)$ and a single function $n_{\rm mag}(H,T)$. For the 
latter we proceed, as shown in Fig.~\ref{fig:AFM_MvT}(b), by 
comparing our 2D spin-wave estimates at all low temperatures (solid lines) 
and our Weiss-field estimates at all high temperatures (dashed lines) with 
our $m(T)$ data at 1 T and 5 T. As in Fig.~\ref{fig:AFM_MvT}(a), we also use 
our Weiss-field estimates at 9 and 18 T. By inspection, the upper temperature, 
$T_u(H)$, beyond which the Weiss-field approach is quantitatively accurate, 
may be taken as $T_u(H) = 40$ K for all fields; in fact a constant $T_u(H)$ 
would not be anticipated simply from the extent of region (\textbf{II}) in 
Fig.~\ref{fig:5n}, and this result is actually a consequence of the 
additional effects of magnetic correlations above $T_N$ at low fields. 
The lower temperature, $T_l(H)$, below which the 2D spin-wave result is 
quantitatively accurate, appears to be approximately 25, 30, and 35 K 
at 5, 9, and 18 T, values with do track the edge of region (\textbf{I}) as 
represented in Fig.~\ref{fig:5n}. At 1 T, the spin-wave approach is clearly 
no longer appropriate, because the tendency to AF order created by $J'$ is 
not fully suppressed (Fig.~\ref{fig:5n}; while we may use the $m(T)$ data 
here for quantitative purposes, we observe that the spin-wave result for a 
monolayer appears to be reliable to approximately 20 K. We note that these 
values of $T_l(H)$ are not dissimilar to the values $T_{\rm max}(H)$ obtained 
from the peak in the magnetic specific heat (Sec.~\ref{sexp}C), and comment 
that this latter scale might indeed be anticipated as the upper limit to an 
ordered magnetic configuration on which to base a 2D spin-wave treatment. To 
bridge the shrinking gap between $T_l(H)$ and $T_u(H)$, we adopt the spline 
fits shown in Fig.~\ref{fig:AFM_MvT}(b).

For $\lambda(T)$, as noted in Sec.~\ref{smodel}B, we have tested three 
functions that reproduce the four data points spanning the peak in this 
quantity shown in the inset of Fig.~\ref{fig:wH}; one has a Gaussian form, 
$\lambda_1 (T)= a_1 \exp [-(T - T_1)^2/ 2 \sigma^2]$, one an exponential form, 
$\lambda_2 (T) = a_2 T^b \exp (-T/T_2)$, and one a Lorentzian form, $\lambda_3 
(T) = a_3 \gamma/\pi [(T - T_3)^2 + \gamma^2]$. The three functions differ only 
in the rate at which $\lambda$ vanishes in the high-$T$ regime and all three 
may be used to obtain consistent descriptions of the data with only minor 
differences in the value of the estimated $\kph(T)$ at $T > 40$ K and in the 
temperature at which all datasets converge. Because of a mild but unexpected 
bulge around 40 K in our measured $\ka(T)$ data (Fig.~\ref{fig:kT}), which is 
most pronounced at 18 T and which we believe is not intrinsic to the phonon 
thermal conductivity, a rapid vanishing of $\lambda(T)$ cannot accommodate 
this feature. To avoid an ill-defined subtraction of this poorly understood 
contribution, it is most convenient to use the Lorentzian continuation of 
$\lambda(T)$, but we make no claim to have proven a physical underpinning 
for this form. 

Inverting our $\ka(H,T)$ data using these estimates of $n_{\rm mag}(H,T)$ and 
$\lambda(T)$ provides four different curves for $\kph^{\rm est}(T)$ based on our 
measurements at 1, 5, 9, and 18 T. From Fig.~\ref{fig:KvTph} it is clear that 
all four estimates of $\kph(T)$ are very similar across the full range of 
temperatures and that they converge completely in both the low- and high-$T$ 
limits. In detail, the estimate based on our 1 T data, which are affected by 
the AF ordering tendencies, are something of an outlier. Otherwise, the 
three estimates based on 5, 9, and 18 T converge to high accuracy at all 
temperatures, with a maximum deviation of order 10\% at $T_p$. Because the 
minor bulge in our $\ka(T)$ data around 40 K (Fig.~\ref{fig:kT}) is strongest 
at 18 T, we judge the near-identical $\kph(T)$ curves deduced from $\mu_0 H
 = 5$ and 9 T to be the most representative and adopt these as our definitive 
result for the phonon thermal conductivity.

\begin{figure}[t]
\includegraphics[width=\linewidth]{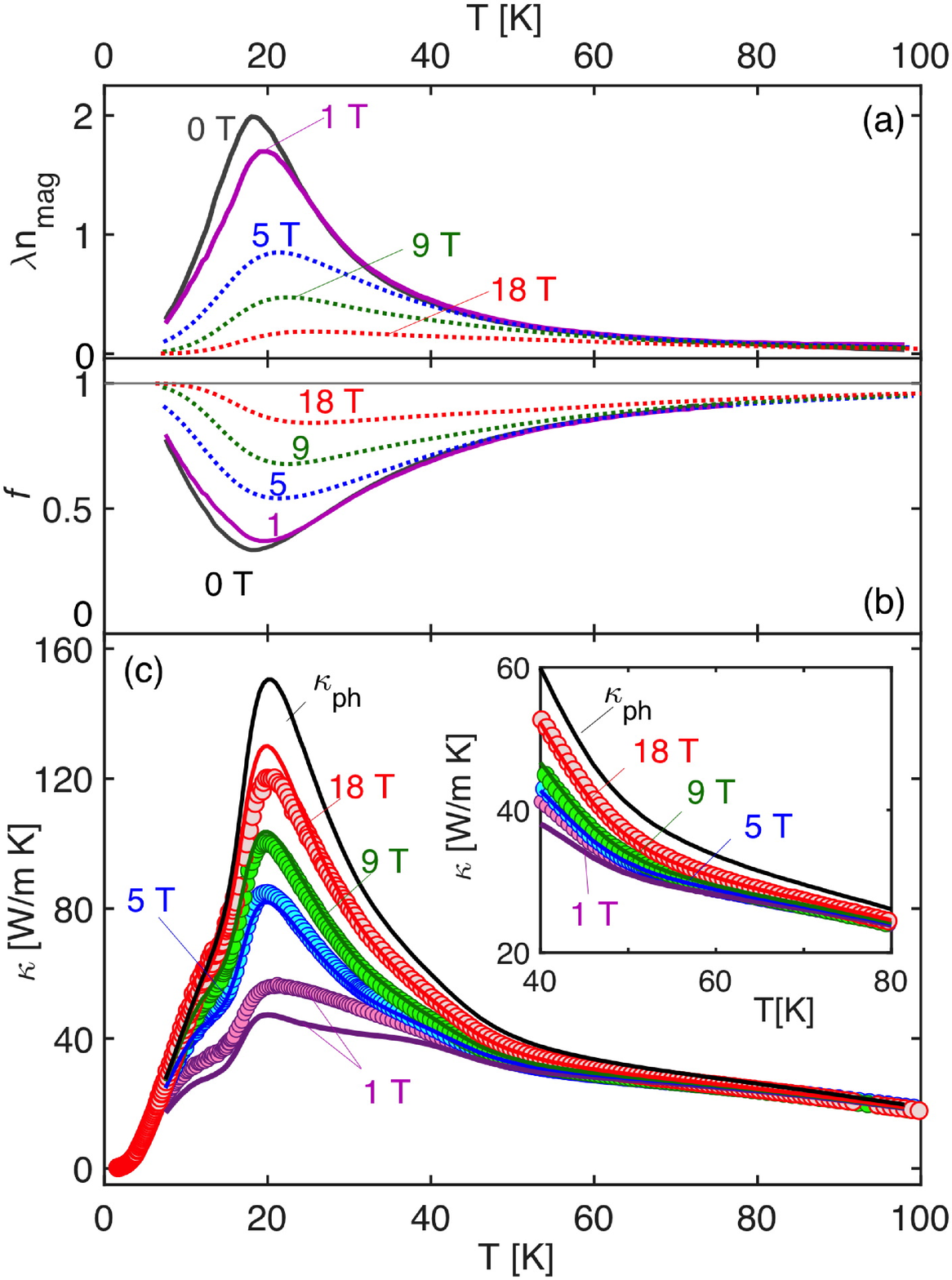}
\caption{(a) The quantity $\lambda \nmag$ shown as a function of temperature 
for fields of 0, 1, 5, 9, and 18 T. Dotted lines for 5, 9, and 18 T are 
obtained from our estimated $\nmag(H,T)$ (Secs.~\ref{smodel}A and 
\ref{smodel}B) multiplied by the Lorentzian continuation of $\lambda(T)$
(see text); solid lines for 0 and 1 T, shown for comparison, were 
obtained directly from the data. (b) Suppression factor, $f = 1/(1 + \lambda 
\nmag)$, corresponding to $\lambda \nmag$ in panel (a). (c) Reconstruction of 
our measured $\ka(H,T)$ data, shown as discrete points, based on the empirical 
model of Eq.~(\ref{eq:wxx}). Calculated $\ka(H,T)$ values (solid lines) are 
based on a single curve for $\kph(T)$ (solid black line) and use the Lorentzian 
continuation of $\lambda(T)$. Inset: magnification of the high-$T$ regime 
($T > 40$ K) to illustrate convergence of the data and calculations to 
$\kph(T)$.}
\label{fig:Krecon}
\end{figure}

Fixing $\kph(T)$ allows us unprecedented physical insight into the effects 
of phonon scattering by the spin fluctuations, and into the way in which these 
effects are suppressed by the magnetic field. In Fig.~\ref{fig:Krecon}(a) we 
show the quantity $\lambda \nmag$, which is equivalent to the relative 
scattering rate $\tau_0/\tau_{\rm mag}$. At 5, 9, and 18 T (dotted lines), 
$\lambda \nmag$ displays a clear peak around 20 K that is suppressed from 
a value of nearly 1 by a factor of approximately 2 at each field step. For 
comparison, we show as solid lines the equivalent quantities for 0 and 1 T, 
which we have taken directly from our data because our methods for estimating 
$\nmag$ are not suitable at these low fields. We observe the same general 
form, but with the peak twice as strong again. 

In Fig.~\ref{fig:Krecon}(b) we show the suppression factor, $f = (1 + \lambda 
\nmag)^{-1}$, corresponding to Fig.~\ref{fig:Krecon}(a), which is an inverted 
function reaching a minimum of 1/3 at $T_p$ for a field of 0 T, where $\lambda 
\nmag \simeq 2$. Thus the quantitative conclusion concerning spin-fluctuation 
scattering of the phonon modes in \crcl~is that this mechanism is so effective 
around the N\'eel temperature of the AF ordered phase that 2/3 of the phonon 
contributions to the thermal conductivity are removed. The spin-scattering 
effect is not at all resonant, retaining a significant value over the entire
range of temperatures from 0 to 100 K. Finally, in Fig.~\ref{fig:Krecon}(c) 
we use our deduced values of $\kph(T)$, $\nmag(H,T)$, and $\lambda(T)$ to 
reconstruct our measured $\ka(H,T)$ data for all fields and temperatures. 
Quantitatively excellent agreement is achieved in all cases, 
with deviations from 1\% at 5 and 9 T to 15\% around the peak at 1 T, where 
in any case our approximations are of limited validity. We stress that this 
is not a circular exercise, because all of the field-dependence of $\ka(H,T)$ 
is reproduced using only $\nmag(H,T)$. We comment also that $\kph(T)$ does 
lie significantly above our 18 T data; while this separation may be slightly 
exaggerated in the high-$T$ regime [inset Fig.~\ref{fig:Krecon}(c)] by our 
use of the Lorentzian continuation, the 20\% difference around the peak 
position is a robust result that serves as a warning against assuming that 
high-field data must lie beyond the range of spin fluctuations. 

It is clear that our models provide a quantitatively accurate fit to the 
thermal conductivity at all fields and temperatures with only one exception. 
This is the low-field plateau in the measured $\ka(T)$ that arises due to the 
ordering transition, a feature that lies explicitly beyond our modelling. 
Thus we have shown that \crcl~provides an ideal system for modelling the 
magnetic fluctuations at finite fields and temperatures, and that the 
dramatic suppression of its thermal conductivity caused by these fluctuations 
can further be modelled by multiplying their density by a field-independent 
phonon scattering strength.

\section{Discussion and Conclusion} 
\label{sdc} 

We first reiterate what we have achieved in describing the thermal conductivity 
of a correlated magnetic insulator with significant spin-lattice coupling. We 
have shown that, beyond a very narrow regime of field and temperature hosting 
3D magnetic order, the thermal conductivity is due entirely to phonons, but 
that the contributions of these phonons is subject to a suppression factor. 
This suppression is due entirely to scattering from spin fluctuations that 
are not coherent in 3D, and can be extremely strong (up to 65\%), but can 
itself be suppressed to zero by a sufficiently strong magnetic field. This 
phenomenon can be captured quantitatively by expressing the suppression in 
terms of only two parameters, an independently determined number of active 
magnetic fluctuations and a dimensionless parameter for their scattering 
efficiency. This latter turns out to be entirely independent of the field, 
meaning that the scattering strength is dictated only by the temperature, 
while the full suppressive effects of the applied field are contained within 
the number of fluctuations. 

Applying this very general framework, for \crcl~we model the 
fluctuation number, $n_{\rm mag}(H,T)$, by considering the two regimes of high 
and low $H/T$. In the former, magnetic fluctuations take the form of 
field-gapped 2D spin waves in the FM honeycomb planes and $n_{\rm mag}$ is 
given by a conventional spin-wave theory, but the complete lack of 
interlayer coherence means that these modes are not only incapable of 
transporting heat themselves but are destructive to the more 3D phonons 
that do so. In the latter, the system is paramagnetic and dominated by 
thermal fluctuations, but it is still field-polarized and thus the 
intrinsic intralayer FM interaction continues to play a role. The energy 
scale of this interaction, of order 15 K, is in fact fundamental to the 
thermodynamic and transport response of the system at all fields and 
temperatures, and its fingerprints are found in quantities ranging from 
the magnetic specific heat to the empirically determined phonon 
scattering-strength parameter, $\lambda(T)$, which peaks in temperature 
around this value before falling to zero. With these ingredients, our 
formalism can reproduce, and indeed predict the form of, $\kappa(T)$ at 
different applied fields over the entire temperature range. Discrepancies 
between the data and the model appear only at low fields in the vicinity of 
$T_N$, where $\nmag$ cannot describe adequately the population of fluctuating 
spins in the critical regime.

A key strength of our modelling procedure is its basis in the simple
phonon-scattering form expressed in Eq.~(\ref{eq:wxx}). Despite its  
phenomenological nature, our approach is both completely general and 
fully quantitative. It is not dependent on specific properties of the 
magnetic state of the system and, crucially, it is independent of the 
nature of the magnetic excitations, which in the quantum limit may not 
necessarily be conventional magnons, but topological ones or even only 
fractions of a single spin-flip \cite{Sologubenko2000,Sologubenko2001}.
Our treatment is also insensitive to specific phonon scattering mechanisms, 
which as Sec.~\ref{sintro} makes clear can take many possible forms; in this 
context we note again that the spin-fluctuation scattering we consider is 
non-resonant and that these fluctuations are paramagnons (quasi-FM spin 
fluctuations in the paramagnetic regime) over the entire range of $T$ and 
$H$. While we have kept our treatment independent of assumptions about 
the phonon thermal transport in order to make it fully quantitative 
(Sec.~\ref{smodel}C), one may also ask whether the analysis could be 
improved, or otherwise brought into contact with conventional treatments 
based on the specific heat or the DC framework. While this is certainly 
possible in parts of the parameter space, we have not been able to find 
a global description for \crcl~within either approach, and on this basis 
would not expect this type of traditional formalism to be suitable for 
general magnetic materials. We stress again that, if the $\kph(T)$ curve 
we extract from our data is regarded as a given, the number of ``free'' 
parameters in our modelling procedure can be argued to be zero, should one 
allow that ${\tilde J}$ in region (\textbf{I}) and $B_{\rm mol}$ in region 
(\textbf{II}) are given by the in-plane FM energy scale, $3 J \approx T_N$, 
and that the form of the high-$T$ vanishing of $\lambda(T)$ is immaterial.

An essential aspect of our study is the issue of system dimensionality. 
While all magnetic insulators are 3D, in low-dimensional quantum magnets 
the regime of 3D behavior may be a very small part of the ($H,T$) parameter 
space. In \crcl, the stronger coupling in the honeycomb layers mandates a 
2D treatment in the regime of large $H/T$, where the field destroys 3D 
correlations but does not damage the 2D FM correlations; by contrast, 
thermal fluctuations act to damage all correlations. The FM nature of the 
layers also has another unexpected consequence in that, although the CrX$_3$ 
systems are regarded structurally as van der Waals materials, featuring a 
very low cohesive energy for exfoliation \cite{McGuire2017}, \crcl~does 
not show the conventional features of a 2D magnet. In the specific heat, 
the 3D ordering peak effectively coincides with the broad maximum 
characterizing the majority of the spin-fluctuation processes, while the 
minimal anisotropy is also consistent with 3D magnetism. In more detail, 
the interlayer superexchange interaction, $J' \approx 1$ K, while not an 
insignificant fraction of the intralayer $J = - 5.25$ K, is indeed rather 
smaller, and it is the FM nature of the in-plane order that allows $J'$ 
to ``leverage'' a $T_N$ scale of order $3|J|$. We comment that layered 
magnetic materials are ubiquitous both in condensed matter and in the 
heterostructures being fabricated for spintronic functionalities, and hence 
our considerations can be expected to have far-reaching applicability. 

While it is intuitively clear that the origin of the giant magnetoconductivity 
we observe lies in ``strong spin-lattice coupling'' \cite{McGuire2017}, we 
stress that this is not merely another spin-phonon story. The qualitative 
difference in the present study is that we are measuring a transport property, 
meaning a property exclusively of the excitations in the spin and lattice 
sectors of the system. In this sense our focus is a specific and sensitive
probe of a much less commonly studied aspect of magnetoelastic coupling, 
namely the nature and scattering of these two sets of excitations over the 
complete ($H,T$) parameter space. 

One may nonetheless ask why, quantitatively, the scattering effect is so 
strong in \crcl. Here we point to the possibility that the spin-phonon 
coupling can be relatively normal but $J$ is in fact anomalously small. As 
noted in Sec.~\ref{smm}, the FM $J$ is a consequence of the near-90$^\circ$ 
Cr-Cl-Cr bond angle enforced by edge-sharing CrCl$_6$ octahedra, and this 
type of interaction is far smaller in magnitude than comparable AF bonds at 
higher angles. In general, the interaction strength is very sensitive to the 
bond angle in this regime, implying that small phononic displacements may 
have strong relative effects; when normalized by the small $J$ values, 
these magnetoelastic effects then appear in the conventional range. While 
first-principles calculations have been performed recently to accompany the 
experimental observation that the interlayer magnetic interaction strength, 
$J'$, changes over a wide range in few-layer \crcl~samples \cite{kleinetal}, 
we are not aware of calculations investigating the lattice-sensitivity of the 
in-plane interaction, $J$, which could in principle be modulated by pressure 
in bulk samples or by substrate choice in few-layer heterostructured samples. 

Returning to $\alpha$-RuCl$_3$, small Heisenberg interactions, $J$, are also a 
sought-after feature of candidate Kitaev materials. When these dominant 
magnetic effects are suppressed, the remaining terms in the anisotropic 
spin Hamiltonian are thought to give rise to fractional spin excitations (of 
Majorana \cite{Kitaev2006,Banerjee2016,Banerjee2017} or generalized Majorana 
character \cite{rwnl}) and to spin-liquid ground states in the presence of 
an applied magnetic field \cite{Leahy2017,Zheng2017,rln}. It is clear 
\cite{Leahy2017} that the thermal conductivity measured in $\alpha$-RuCl$_3$ 
shows strong spin-phonon scattering effects over the entire range of 
temperatures, but a detailed interpretation lies beyond the scope of a DC 
approach \cite{Hentrich2018} and a theoretical analysis of phonon scattering 
by fractionalized spins is still awaited. We comment in passing that phonon 
coupling to chiral Majorana edge states has recently been invoked \cite{ryhsb,
rvr} as an ingredient essential for the interpretation of controversial 
thermal Hall conductivity data reported \cite{rkbs} for $\alpha$-RuCl$_3$ 
at finite fields; however, we stress that the strong spin-phonon scattering 
observed in both $\alpha$-RuCl$_3$ and \crcl, and which we model here, involves 
the bulk spin excitations. We suggest that the more general, data-oriented 
approach we adopt for \crcl~may help to clarify the situation even in the 
absence of a microscopic discussion of phonon scattering by fractional spin 
excitations. 

Returning to the higher chromium trihalides, CrBr$_3$ has been proposed 
\cite{Pershoguba2018} as a candidate for hosting topological magnons. CrI$_3$ 
is known \cite{Huang2017} to present a situation where the bulk material has 
FM interlayer interactions, but the few-layer form takes on a different 
interlayer structure and these interactions become AF. In a similar vein, it 
has been found very recently in \crcl~that the interlayer interaction remains 
AF \cite{caoetal} as the system thickness is reduced to two layers, while 
showing the dramatic increase noted above \cite{kleinetal}. Efforts to 
include CrBr$_3$ in a systematic comparison are ongoing \cite{kimetal}. These 
results have drawn a great deal of attention with a view to fabricating highly 
controllable spintronic materials, possibly functioning with topologically 
protected information. While thermal conductivity measurements on few- or 
many-layer samples are not yet available, our results offer both a general 
framework for analyzing the different possible contributions to spin and 
thermal transport and a general warning concerning the need to take full 
account of spin-phonon scattering effects. 

In summary, we have investigated the thermal conductivity of the layered 
ferromagnet \crcl~over a wide range of temperature and magnetic field. We 
find a giant field-induced enhancement of the phononic contribution at all 
temperatures below 70 K, pointing towards a strong spin-fluctuation scattering 
effect. We construct an empirical model for the thermal conductivity by 
introducing a general framework based on two quantities that can be 
determined separately, the number of active spin-flip processes and their  
efficiency in scattering phonons. This formalism provides a quantitative 
description of our measured data at all fields and temperatures, has 
predictive power in unmeasured regions, and allows an accurate extraction 
of the purely phononic response. We anticipate that this approach will find 
wide application in interpreting the spin and thermal transport properties 
of many insulating magnetic materials, where spin-phonon scattering is a 
strong and unavoidable feature of the physics. 

\begin{acknowledgements}

CAP and IAL acknowledge the support of the Colorado Energy Research 
Collaboration. A portion of this work was performed at the National 
High Magnetic Field Laboratory, which is supported by National Science 
Foundation Cooperative Agreement No.~DMR-1644779 and by the State of 
Florida. Work at CAU was supported through National Research Foundation 
Grant No.~2019R1A2C3006189 by the Ministry of Science and ICT of South 
Korea.

\end{acknowledgements}

%


\begin{thebibliography}{49}%
\makeatletter
\providecommand \@ifxundefined [1]{%
 \@ifx{#1\undefined}
}%
\providecommand \@ifnum [1]{%
 \ifnum #1\expandafter \@firstoftwo
 \else \expandafter \@secondoftwo
 \fi
}%
\providecommand \@ifx [1]{%
 \ifx #1\expandafter \@firstoftwo
 \else \expandafter \@secondoftwo
 \fi
}%
\providecommand \natexlab [1]{#1}%
\providecommand \enquote  [1]{``#1''}%
\providecommand \bibnamefont  [1]{#1}%
\providecommand \bibfnamefont [1]{#1}%
\providecommand \citenamefont [1]{#1}%
\providecommand \href@noop [0]{\@secondoftwo}%
\providecommand \href [0]{\begingroup \@sanitize@url \@href}%
\providecommand \@href[1]{\@@startlink{#1}\@@href}%
\providecommand \@@href[1]{\endgroup#1\@@endlink}%
\providecommand \@sanitize@url [0]{\catcode `\\12\catcode `\$12\catcode
  `\&12\catcode `\#12\catcode `\^12\catcode `\_12\catcode `\%12\relax}%
\providecommand \@@startlink[1]{}%
\providecommand \@@endlink[0]{}%
\providecommand \url  [0]{\begingroup\@sanitize@url \@url }%
\providecommand \@url [1]{\endgroup\@href {#1}{\urlprefix }}%
\providecommand \urlprefix  [0]{URL }%
\providecommand \Eprint [0]{\href }%
\providecommand \doibase [0]{http://dx.doi.org/}%
\providecommand \selectlanguage [0]{\@gobble}%
\providecommand \bibinfo  [0]{\@secondoftwo}%
\providecommand \bibfield  [0]{\@secondoftwo}%
\providecommand \translation [1]{[#1]}%
\providecommand \BibitemOpen [0]{}%
\providecommand \bibitemStop [0]{}%
\providecommand \bibitemNoStop [0]{.\EOS\space}%
\providecommand \EOS [0]{\spacefactor3000\relax}%
\providecommand \BibitemShut  [1]{\csname bibitem#1\endcsname}%
\let\auto@bib@innerbib\@empty
\bibitem [{\citenamefont {Ziman}(1960)}]{Zimanbook}%
  \BibitemOpen
  \bibfield  {author} {\bibinfo {author} {\bibfnamefont {J.~M.}\ \bibnamefont
  {Ziman}},\ }\href@noop {} {\emph {\bibinfo {title} {Electrons and Phonons}}}\
  (\bibinfo  {publisher} {Oxford University Press, Oxford},\ \bibinfo {year}
  {1960})\BibitemShut {NoStop}%
\bibitem [{\citenamefont {Ando}\ \emph {et~al.}(1998)\citenamefont {Ando},
  \citenamefont {Takeya}, \citenamefont {Sisson}, \citenamefont {Doettinger},
  \citenamefont {Tanaka}, \citenamefont {Feigelson},\ and\ \citenamefont
  {Kapitulnik}}]{Ando1998}%
  \BibitemOpen
  \bibfield  {author} {\bibinfo {author} {\bibfnamefont {Y.}~\bibnamefont
  {Ando}}, \bibinfo {author} {\bibfnamefont {J.}~\bibnamefont {Takeya}},
  \bibinfo {author} {\bibfnamefont {D.~L.}\ \bibnamefont {Sisson}}, \bibinfo
  {author} {\bibfnamefont {S.~G.}\ \bibnamefont {Doettinger}}, \bibinfo
  {author} {\bibfnamefont {I.}~\bibnamefont {Tanaka}}, \bibinfo {author}
  {\bibfnamefont {R.~S.}\ \bibnamefont {Feigelson}}, \ and\ \bibinfo {author}
  {\bibfnamefont {A.}~\bibnamefont {Kapitulnik}},\ }\href@noop {} {\bibfield
  {journal} {\bibinfo  {journal} {Phys. Rev. B}\ }\textbf {\bibinfo {volume}
  {58}},\ \bibinfo {pages} {2913} (\bibinfo {year} {1998})}\BibitemShut
  {NoStop}%
\bibitem [{\citenamefont {Sologubenko}\ \emph {et~al.}(2000)\citenamefont
  {Sologubenko}, \citenamefont {Giann\'o}, \citenamefont {Ott}, \citenamefont
  {Ammerahl},\ and\ \citenamefont {Revcolevschi}}]{Sologubenko2000}%
  \BibitemOpen
  \bibfield  {author} {\bibinfo {author} {\bibfnamefont {A.~V.}\ \bibnamefont
  {Sologubenko}}, \bibinfo {author} {\bibfnamefont {K.}~\bibnamefont
  {Giann\'o}}, \bibinfo {author} {\bibfnamefont {H.~R.}\ \bibnamefont {Ott}},
  \bibinfo {author} {\bibfnamefont {U.}~\bibnamefont {Ammerahl}}, \ and\
  \bibinfo {author} {\bibfnamefont {A.}~\bibnamefont {Revcolevschi}},\
  }\href@noop {} {\bibfield  {journal} {\bibinfo  {journal} {Phys. Rev. Lett.}\
  }\textbf {\bibinfo {volume} {84}},\ \bibinfo {pages} {2714} (\bibinfo {year}
  {2000})}\BibitemShut {NoStop}%
\bibitem [{\citenamefont {Hofmann}\ \emph {et~al.}(2001)\citenamefont
  {Hofmann}, \citenamefont {Lorenz}, \citenamefont {Uhrig}, \citenamefont
  {Kierspel}, \citenamefont {Zabara}, \citenamefont {Freimuth}, \citenamefont
  {Kageyama},\ and\ \citenamefont {Ueda}}]{Hofmann2001}%
  \BibitemOpen
  \bibfield  {author} {\bibinfo {author} {\bibfnamefont {M.}~\bibnamefont
  {Hofmann}}, \bibinfo {author} {\bibfnamefont {T.}~\bibnamefont {Lorenz}},
  \bibinfo {author} {\bibfnamefont {G.~S.}\ \bibnamefont {Uhrig}}, \bibinfo
  {author} {\bibfnamefont {H.}~\bibnamefont {Kierspel}}, \bibinfo {author}
  {\bibfnamefont {O.}~\bibnamefont {Zabara}}, \bibinfo {author} {\bibfnamefont
  {A.}~\bibnamefont {Freimuth}}, \bibinfo {author} {\bibfnamefont
  {H.}~\bibnamefont {Kageyama}}, \ and\ \bibinfo {author} {\bibfnamefont
  {Y.}~\bibnamefont {Ueda}},\ }\href@noop {} {\bibfield  {journal} {\bibinfo
  {journal} {Phys. Rev. Lett.}\ }\textbf {\bibinfo {volume} {87}},\ \bibinfo
  {pages} {047202} (\bibinfo {year} {2001})}\BibitemShut {NoStop}%
\bibitem [{\citenamefont {Sales}\ \emph {et~al.}(2002)\citenamefont {Sales},
  \citenamefont {Lumsden}, \citenamefont {Nagler}, \citenamefont {Mandrus},\
  and\ \citenamefont {Jin}}]{Sales2002}%
  \BibitemOpen
  \bibfield  {author} {\bibinfo {author} {\bibfnamefont {B.~C.}\ \bibnamefont
  {Sales}}, \bibinfo {author} {\bibfnamefont {M.~D.}\ \bibnamefont {Lumsden}},
  \bibinfo {author} {\bibfnamefont {S.~E.}\ \bibnamefont {Nagler}}, \bibinfo
  {author} {\bibfnamefont {D.}~\bibnamefont {Mandrus}}, \ and\ \bibinfo
  {author} {\bibfnamefont {R.}~\bibnamefont {Jin}},\ }\href@noop {} {\bibfield
  {journal} {\bibinfo  {journal} {Phys. Rev. Lett.}\ }\textbf {\bibinfo
  {volume} {88}},\ \bibinfo {pages} {095901} (\bibinfo {year}
  {2002})}\BibitemShut {NoStop}%
\bibitem [{\citenamefont {Jin}\ \emph {et~al.}(2003)\citenamefont {Jin},
  \citenamefont {Onose}, \citenamefont {Tokura}, \citenamefont {Mandrus},
  \citenamefont {Dai},\ and\ \citenamefont {Sales}}]{Jin2003}%
  \BibitemOpen
  \bibfield  {author} {\bibinfo {author} {\bibfnamefont {R.}~\bibnamefont
  {Jin}}, \bibinfo {author} {\bibfnamefont {Y.}~\bibnamefont {Onose}}, \bibinfo
  {author} {\bibfnamefont {Y.}~\bibnamefont {Tokura}}, \bibinfo {author}
  {\bibfnamefont {D.}~\bibnamefont {Mandrus}}, \bibinfo {author} {\bibfnamefont
  {P.}~\bibnamefont {Dai}}, \ and\ \bibinfo {author} {\bibfnamefont {B.~C.}\
  \bibnamefont {Sales}},\ }\href@noop {} {\bibfield  {journal} {\bibinfo
  {journal} {Phys. Rev. Lett.}\ }\textbf {\bibinfo {volume} {91}},\ \bibinfo
  {pages} {146601} (\bibinfo {year} {2003})}\BibitemShut {NoStop}%
\bibitem [{\citenamefont {Li}\ \emph {et~al.}(2005)\citenamefont {Li},
  \citenamefont {Taillefer}, \citenamefont {Wang},\ and\ \citenamefont
  {Chen}}]{SYLi2005}%
  \BibitemOpen
  \bibfield  {author} {\bibinfo {author} {\bibfnamefont {S.~Y.}\ \bibnamefont
  {Li}}, \bibinfo {author} {\bibfnamefont {L.}~\bibnamefont {Taillefer}},
  \bibinfo {author} {\bibfnamefont {C.~H.}\ \bibnamefont {Wang}}, \ and\
  \bibinfo {author} {\bibfnamefont {X.~H.}\ \bibnamefont {Chen}},\ }\href@noop
  {} {\bibfield  {journal} {\bibinfo  {journal} {Phys. Rev. Lett.}\ }\textbf
  {\bibinfo {volume} {95}},\ \bibinfo {pages} {156603} (\bibinfo {year}
  {2005})}\BibitemShut {NoStop}%
\bibitem [{\citenamefont {Wu}\ \emph {et~al.}(2016)\citenamefont {Wu},
  \citenamefont {Song}, \citenamefont {Zhao}, \citenamefont {Shi},
  \citenamefont {Xu}, \citenamefont {Zhao}, \citenamefont {Liu}, \citenamefont
  {Zhao},\ and\ \citenamefont {Sun}}]{Wu2015}%
  \BibitemOpen
  \bibfield  {author} {\bibinfo {author} {\bibfnamefont {J.~C.}\ \bibnamefont
  {Wu}}, \bibinfo {author} {\bibfnamefont {J.~D.}\ \bibnamefont {Song}},
  \bibinfo {author} {\bibfnamefont {Z.~Y.}\ \bibnamefont {Zhao}}, \bibinfo
  {author} {\bibfnamefont {J.}~\bibnamefont {Shi}}, \bibinfo {author}
  {\bibfnamefont {H.~S.}\ \bibnamefont {Xu}}, \bibinfo {author} {\bibfnamefont
  {J.~Y.}\ \bibnamefont {Zhao}}, \bibinfo {author} {\bibfnamefont {X.~G.}\
  \bibnamefont {Liu}}, \bibinfo {author} {\bibfnamefont {X.}~\bibnamefont
  {Zhao}}, \ and\ \bibinfo {author} {\bibfnamefont {X.~F.}\ \bibnamefont
  {Sun}},\ }\href@noop {} {\bibfield  {journal} {\bibinfo  {journal} {J. Phys.:
  Condens. Matter}\ }\textbf {\bibinfo {volume} {28}},\ \bibinfo {pages}
  {056002} (\bibinfo {year} {2016})}\BibitemShut {NoStop}%
\bibitem [{\citenamefont {Leahy}\ \emph {et~al.}(2017)\citenamefont {Leahy},
  \citenamefont {Pocs}, \citenamefont {Siegfried}, \citenamefont {Graf},
  \citenamefont {Do}, \citenamefont {Choi}, \citenamefont {Normand},\ and\
  \citenamefont {Lee}}]{Leahy2017}%
  \BibitemOpen
  \bibfield  {author} {\bibinfo {author} {\bibfnamefont {I.~A.}\ \bibnamefont
  {Leahy}}, \bibinfo {author} {\bibfnamefont {C.~A.}\ \bibnamefont {Pocs}},
  \bibinfo {author} {\bibfnamefont {P.~E.}\ \bibnamefont {Siegfried}}, \bibinfo
  {author} {\bibfnamefont {D.}~\bibnamefont {Graf}}, \bibinfo {author}
  {\bibfnamefont {S.-H.}\ \bibnamefont {Do}}, \bibinfo {author} {\bibfnamefont
  {K.-Y.}\ \bibnamefont {Choi}}, \bibinfo {author} {\bibfnamefont
  {B.}~\bibnamefont {Normand}}, \ and\ \bibinfo {author} {\bibfnamefont
  {M.}~\bibnamefont {Lee}},\ }\href@noop {} {\bibfield  {journal} {\bibinfo
  {journal} {Phys. Rev. Lett.}\ }\textbf {\bibinfo {volume} {118}},\ \bibinfo
  {pages} {187203} (\bibinfo {year} {2017})}\BibitemShut {NoStop}%
\bibitem [{\citenamefont {Fletcher}\ \emph {et~al.}(1967)\citenamefont
  {Fletcher}, \citenamefont {Gardner}, \citenamefont {Fox},\ and\ \citenamefont
  {Topping}}]{rfgft}%
  \BibitemOpen
  \bibfield  {author} {\bibinfo {author} {\bibfnamefont {J.~M.}\ \bibnamefont
  {Fletcher}}, \bibinfo {author} {\bibfnamefont {W.~E.}\ \bibnamefont
  {Gardner}}, \bibinfo {author} {\bibfnamefont {A.~C.}\ \bibnamefont {Fox}}, \
  and\ \bibinfo {author} {\bibfnamefont {G.}~\bibnamefont {Topping}},\
  }\href@noop {} {\bibfield  {journal} {\bibinfo  {journal} {J. Chem. Soc. A}\
  ,\ \bibinfo {pages} {1038}} (\bibinfo {year} {1967})}\BibitemShut {NoStop}%
\bibitem [{\citenamefont {Plumb}\ \emph {et~al.}(2014)\citenamefont {Plumb},
  \citenamefont {Clancy}, \citenamefont {Sandilands}, \citenamefont {Shankar},
  \citenamefont {Hu}, \citenamefont {Burch}, \citenamefont {Kee},\ and\
  \citenamefont {Kim}}]{rpea}%
  \BibitemOpen
  \bibfield  {author} {\bibinfo {author} {\bibfnamefont {K.~W.}\ \bibnamefont
  {Plumb}}, \bibinfo {author} {\bibfnamefont {J.~P.}\ \bibnamefont {Clancy}},
  \bibinfo {author} {\bibfnamefont {L.~J.}\ \bibnamefont {Sandilands}},
  \bibinfo {author} {\bibfnamefont {V.~V.}\ \bibnamefont {Shankar}}, \bibinfo
  {author} {\bibfnamefont {Y.~F.}\ \bibnamefont {Hu}}, \bibinfo {author}
  {\bibfnamefont {K.~S.}\ \bibnamefont {Burch}}, \bibinfo {author}
  {\bibfnamefont {H.-Y.}\ \bibnamefont {Kee}}, \ and\ \bibinfo {author}
  {\bibfnamefont {Y.-J.}\ \bibnamefont {Kim}},\ }\href@noop {} {\bibfield
  {journal} {\bibinfo  {journal} {Phys. Rev. B}\ }\textbf {\bibinfo {volume}
  {90}},\ \bibinfo {pages} {041112(R)} (\bibinfo {year} {2014})}\BibitemShut
  {NoStop}%
\bibitem [{\citenamefont {Sears}\ \emph {et~al.}(2015)\citenamefont {Sears},
  \citenamefont {Songvilay}, \citenamefont {Plumb}, \citenamefont {Clancy},
  \citenamefont {Qiu}, \citenamefont {Zhao}, \citenamefont {Parshall},\ and\
  \citenamefont {Kim}}]{rsea}%
  \BibitemOpen
  \bibfield  {author} {\bibinfo {author} {\bibfnamefont {J.~A.}\ \bibnamefont
  {Sears}}, \bibinfo {author} {\bibfnamefont {M.}~\bibnamefont {Songvilay}},
  \bibinfo {author} {\bibfnamefont {K.~W.}\ \bibnamefont {Plumb}}, \bibinfo
  {author} {\bibfnamefont {J.~P.}\ \bibnamefont {Clancy}}, \bibinfo {author}
  {\bibfnamefont {Y.}~\bibnamefont {Qiu}}, \bibinfo {author} {\bibfnamefont
  {Y.}~\bibnamefont {Zhao}}, \bibinfo {author} {\bibfnamefont {D.}~\bibnamefont
  {Parshall}}, \ and\ \bibinfo {author} {\bibfnamefont {Y.-J.}\ \bibnamefont
  {Kim}},\ }\href@noop {} {\bibfield  {journal} {\bibinfo  {journal} {Phys.
  Rev. B}\ }\textbf {\bibinfo {volume} {91}},\ \bibinfo {pages} {144420}
  (\bibinfo {year} {2015})}\BibitemShut {NoStop}%
\bibitem [{\citenamefont {Johnson}\ \emph {et~al.}(2015)\citenamefont
  {Johnson}, \citenamefont {Williams}, \citenamefont {Haghighirad},
  \citenamefont {Singleton}, \citenamefont {Zapf}, \citenamefont {Manuel},
  \citenamefont {Mazin}, \citenamefont {Li}, \citenamefont {Jeschke},
  \citenamefont {Valent\'{\i}},\ and\ \citenamefont {Coldea}}]{rjea}%
  \BibitemOpen
  \bibfield  {author} {\bibinfo {author} {\bibfnamefont {R.~D.}\ \bibnamefont
  {Johnson}}, \bibinfo {author} {\bibfnamefont {S.~C.}\ \bibnamefont
  {Williams}}, \bibinfo {author} {\bibfnamefont {A.~A.}\ \bibnamefont
  {Haghighirad}}, \bibinfo {author} {\bibfnamefont {J.}~\bibnamefont
  {Singleton}}, \bibinfo {author} {\bibfnamefont {V.}~\bibnamefont {Zapf}},
  \bibinfo {author} {\bibfnamefont {P.}~\bibnamefont {Manuel}}, \bibinfo
  {author} {\bibfnamefont {I.~I.}\ \bibnamefont {Mazin}}, \bibinfo {author}
  {\bibfnamefont {Y.}~\bibnamefont {Li}}, \bibinfo {author} {\bibfnamefont
  {H.~O.}\ \bibnamefont {Jeschke}}, \bibinfo {author} {\bibfnamefont
  {R.}~\bibnamefont {Valent\'{\i}}}, \ and\ \bibinfo {author} {\bibfnamefont
  {R.}~\bibnamefont {Coldea}},\ }\href@noop {} {\bibfield  {journal} {\bibinfo
  {journal} {Phys. Rev. B}\ }\textbf {\bibinfo {volume} {92}},\ \bibinfo
  {pages} {235119} (\bibinfo {year} {2015})}\BibitemShut {NoStop}%
\bibitem [{\citenamefont {Cao}\ \emph {et~al.}(2016)\citenamefont {Cao},
  \citenamefont {Banerjee}, \citenamefont {Yan}, \citenamefont {Bridges},
  \citenamefont {Lumsden}, \citenamefont {Mandrus}, \citenamefont {Tennant},
  \citenamefont {Chakoumakos},\ and\ \citenamefont {Nagler}}]{rcaoetal}%
  \BibitemOpen
  \bibfield  {author} {\bibinfo {author} {\bibfnamefont {H.~B.}\ \bibnamefont
  {Cao}}, \bibinfo {author} {\bibfnamefont {A.}~\bibnamefont {Banerjee}},
  \bibinfo {author} {\bibfnamefont {J.-Q.}\ \bibnamefont {Yan}}, \bibinfo
  {author} {\bibfnamefont {C.~A.}\ \bibnamefont {Bridges}}, \bibinfo {author}
  {\bibfnamefont {M.~D.}\ \bibnamefont {Lumsden}}, \bibinfo {author}
  {\bibfnamefont {D.~G.}\ \bibnamefont {Mandrus}}, \bibinfo {author}
  {\bibfnamefont {D.~A.}\ \bibnamefont {Tennant}}, \bibinfo {author}
  {\bibfnamefont {B.~C.}\ \bibnamefont {Chakoumakos}}, \ and\ \bibinfo {author}
  {\bibfnamefont {S.~E.}\ \bibnamefont {Nagler}},\ }\href@noop {} {\bibfield
  {journal} {\bibinfo  {journal} {Phys. Rev. B}\ }\textbf {\bibinfo {volume}
  {93}},\ \bibinfo {pages} {134423} (\bibinfo {year} {2016})}\BibitemShut
  {NoStop}%
\bibitem [{\citenamefont {Banerjee}\ \emph {et~al.}(2016)\citenamefont
  {Banerjee}, \citenamefont {Bridges}, \citenamefont {Yan}, \citenamefont
  {Aczel}, \citenamefont {Li}, \citenamefont {Stone}, \citenamefont {Granroth},
  \citenamefont {Lumsden}, \citenamefont {Yiu}, \citenamefont {Knolle},
  \citenamefont {Bhattacharjee}, \citenamefont {Kovrizhin}, \citenamefont
  {Moessner}, \citenamefont {Tennant}, \citenamefont {Mandrus},\ and\
  \citenamefont {Nagler}}]{Banerjee2016}%
  \BibitemOpen
  \bibfield  {author} {\bibinfo {author} {\bibfnamefont {A.}~\bibnamefont
  {Banerjee}}, \bibinfo {author} {\bibfnamefont {C.~A.}\ \bibnamefont
  {Bridges}}, \bibinfo {author} {\bibfnamefont {J.~Q.}\ \bibnamefont {Yan}},
  \bibinfo {author} {\bibfnamefont {A.~A.}\ \bibnamefont {Aczel}}, \bibinfo
  {author} {\bibfnamefont {L.}~\bibnamefont {Li}}, \bibinfo {author}
  {\bibfnamefont {M.~B.}\ \bibnamefont {Stone}}, \bibinfo {author}
  {\bibfnamefont {G.~E.}\ \bibnamefont {Granroth}}, \bibinfo {author}
  {\bibfnamefont {M.~D.}\ \bibnamefont {Lumsden}}, \bibinfo {author}
  {\bibfnamefont {Y.}~\bibnamefont {Yiu}}, \bibinfo {author} {\bibfnamefont
  {J.}~\bibnamefont {Knolle}}, \bibinfo {author} {\bibfnamefont
  {S.}~\bibnamefont {Bhattacharjee}}, \bibinfo {author} {\bibfnamefont {D.~L.}\
  \bibnamefont {Kovrizhin}}, \bibinfo {author} {\bibfnamefont {R.}~\bibnamefont
  {Moessner}}, \bibinfo {author} {\bibfnamefont {D.~A.}\ \bibnamefont
  {Tennant}}, \bibinfo {author} {\bibfnamefont {D.~G.}\ \bibnamefont
  {Mandrus}}, \ and\ \bibinfo {author} {\bibfnamefont {S.~E.}\ \bibnamefont
  {Nagler}},\ }\href@noop {} {\bibfield  {journal} {\bibinfo  {journal} {Nature
  Mater.}\ }\textbf {\bibinfo {volume} {15}},\ \bibinfo {pages} {733} (\bibinfo
  {year} {2016})}\BibitemShut {NoStop}%
\bibitem [{\citenamefont {Zheng}\ \emph {et~al.}(2017)\citenamefont {Zheng},
  \citenamefont {Ran}, \citenamefont {Li}, \citenamefont {Wang}, \citenamefont
  {Wang}, \citenamefont {Liu}, \citenamefont {Liu}, \citenamefont {Normand},
  \citenamefont {Wen},\ and\ \citenamefont {Yu}}]{Zheng2017}%
  \BibitemOpen
  \bibfield  {author} {\bibinfo {author} {\bibfnamefont {J.}~\bibnamefont
  {Zheng}}, \bibinfo {author} {\bibfnamefont {K.}~\bibnamefont {Ran}}, \bibinfo
  {author} {\bibfnamefont {T.}~\bibnamefont {Li}}, \bibinfo {author}
  {\bibfnamefont {J.}~\bibnamefont {Wang}}, \bibinfo {author} {\bibfnamefont
  {P.-S.}\ \bibnamefont {Wang}}, \bibinfo {author} {\bibfnamefont
  {B.}~\bibnamefont {Liu}}, \bibinfo {author} {\bibfnamefont {Z.-X.}\
  \bibnamefont {Liu}}, \bibinfo {author} {\bibfnamefont {B.}~\bibnamefont
  {Normand}}, \bibinfo {author} {\bibfnamefont {J.}~\bibnamefont {Wen}}, \ and\
  \bibinfo {author} {\bibfnamefont {W.}~\bibnamefont {Yu}},\ }\href@noop {}
  {\bibfield  {journal} {\bibinfo  {journal} {Phys. Rev. Lett.}\ }\textbf
  {\bibinfo {volume} {119}},\ \bibinfo {pages} {227208} (\bibinfo {year}
  {2017})}\BibitemShut {NoStop}%
\bibitem [{\citenamefont {Ponomaryov}\ \emph {et~al.}(2017)\citenamefont
  {Ponomaryov}, \citenamefont {Schulze}, \citenamefont {Wosnitza},
  \citenamefont {Lampen-Kelley}, \citenamefont {Banerjee}, \citenamefont {Yan},
  \citenamefont {Bridges}, \citenamefont {Mandrus}, \citenamefont {Nagler},
  \citenamefont {Kolezhuk},\ and\ \citenamefont {Zvyagin}}]{Ponomaryov2017}%
  \BibitemOpen
  \bibfield  {author} {\bibinfo {author} {\bibfnamefont {A.~N.}\ \bibnamefont
  {Ponomaryov}}, \bibinfo {author} {\bibfnamefont {E.}~\bibnamefont {Schulze}},
  \bibinfo {author} {\bibfnamefont {J.}~\bibnamefont {Wosnitza}}, \bibinfo
  {author} {\bibfnamefont {P.}~\bibnamefont {Lampen-Kelley}}, \bibinfo {author}
  {\bibfnamefont {A.}~\bibnamefont {Banerjee}}, \bibinfo {author}
  {\bibfnamefont {J.~Q.}\ \bibnamefont {Yan}}, \bibinfo {author} {\bibfnamefont
  {C.~A.}\ \bibnamefont {Bridges}}, \bibinfo {author} {\bibfnamefont {D.~G.}\
  \bibnamefont {Mandrus}}, \bibinfo {author} {\bibfnamefont {S.~E.}\
  \bibnamefont {Nagler}}, \bibinfo {author} {\bibfnamefont {A.~K.}\
  \bibnamefont {Kolezhuk}}, \ and\ \bibinfo {author} {\bibfnamefont {S.~A.}\
  \bibnamefont {Zvyagin}},\ }\href@noop {} {\bibfield  {journal} {\bibinfo
  {journal} {Phys. Rev. B}\ }\textbf {\bibinfo {volume} {96}},\ \bibinfo
  {pages} {241107} (\bibinfo {year} {2017})}\BibitemShut {NoStop}%
\bibitem [{\citenamefont {Banerjee}\ \emph {et~al.}(2017)\citenamefont
  {Banerjee}, \citenamefont {Yan}, \citenamefont {Knolle}, \citenamefont
  {Bridges}, \citenamefont {Stone}, \citenamefont {Lumsden}, \citenamefont
  {Mandrus}, \citenamefont {Tennant}, \citenamefont {Moessner},\ and\
  \citenamefont {Nagler}}]{Banerjee2017}%
  \BibitemOpen
  \bibfield  {author} {\bibinfo {author} {\bibfnamefont {A.}~\bibnamefont
  {Banerjee}}, \bibinfo {author} {\bibfnamefont {J.~Q.}\ \bibnamefont {Yan}},
  \bibinfo {author} {\bibfnamefont {J.}~\bibnamefont {Knolle}}, \bibinfo
  {author} {\bibfnamefont {C.~A.}\ \bibnamefont {Bridges}}, \bibinfo {author}
  {\bibfnamefont {M.~B.}\ \bibnamefont {Stone}}, \bibinfo {author}
  {\bibfnamefont {M.~D.}\ \bibnamefont {Lumsden}}, \bibinfo {author}
  {\bibfnamefont {D.~G.}\ \bibnamefont {Mandrus}}, \bibinfo {author}
  {\bibfnamefont {D.~A.}\ \bibnamefont {Tennant}}, \bibinfo {author}
  {\bibfnamefont {R.}~\bibnamefont {Moessner}}, \ and\ \bibinfo {author}
  {\bibfnamefont {S.~E.}\ \bibnamefont {Nagler}},\ }\href@noop {} {\bibfield
  {journal} {\bibinfo  {journal} {Science}\ }\textbf {\bibinfo {volume}
  {356}},\ \bibinfo {pages} {1055} (\bibinfo {year} {2017})}\BibitemShut
  {NoStop}%
\bibitem [{\citenamefont {Banerjee}\ \emph {et~al.}(2018)\citenamefont
  {Banerjee}, \citenamefont {Lampen-Kelley}, \citenamefont {Knolle},
  \citenamefont {Balz}, \citenamefont {Aczel}, \citenamefont {Winn},
  \citenamefont {Liu}, \citenamefont {Pajerowski}, \citenamefont {Yan},
  \citenamefont {Bridges}, \citenamefont {Savici}, \citenamefont {Chakoumakos},
  \citenamefont {Lumsden}, \citenamefont {Tennant}, \citenamefont {Moessner},
  \citenamefont {Mandrus},\ and\ \citenamefont {Nagler}}]{Banerjee2018}%
  \BibitemOpen
  \bibfield  {author} {\bibinfo {author} {\bibfnamefont {A.}~\bibnamefont
  {Banerjee}}, \bibinfo {author} {\bibfnamefont {P.}~\bibnamefont
  {Lampen-Kelley}}, \bibinfo {author} {\bibfnamefont {J.}~\bibnamefont
  {Knolle}}, \bibinfo {author} {\bibfnamefont {C.}~\bibnamefont {Balz}},
  \bibinfo {author} {\bibfnamefont {A.~A.}\ \bibnamefont {Aczel}}, \bibinfo
  {author} {\bibfnamefont {B.}~\bibnamefont {Winn}}, \bibinfo {author}
  {\bibfnamefont {Y.}~\bibnamefont {Liu}}, \bibinfo {author} {\bibfnamefont
  {D.}~\bibnamefont {Pajerowski}}, \bibinfo {author} {\bibfnamefont {J.-Q.}\
  \bibnamefont {Yan}}, \bibinfo {author} {\bibfnamefont {C.~A.}\ \bibnamefont
  {Bridges}}, \bibinfo {author} {\bibfnamefont {A.~T.}\ \bibnamefont {Savici}},
  \bibinfo {author} {\bibfnamefont {B.~C.}\ \bibnamefont {Chakoumakos}},
  \bibinfo {author} {\bibfnamefont {M.~D.}\ \bibnamefont {Lumsden}}, \bibinfo
  {author} {\bibfnamefont {D.~A.}\ \bibnamefont {Tennant}}, \bibinfo {author}
  {\bibfnamefont {R.}~\bibnamefont {Moessner}}, \bibinfo {author}
  {\bibfnamefont {D.~G.}\ \bibnamefont {Mandrus}}, \ and\ \bibinfo {author}
  {\bibfnamefont {S.~E.}\ \bibnamefont {Nagler}},\ }\href@noop {} {\bibfield
  {journal} {\bibinfo  {journal} {npj Quantum Materials}\ }\textbf {\bibinfo
  {volume} {3}},\ \bibinfo {pages} {8} (\bibinfo {year} {2018})}\BibitemShut
  {NoStop}%
\bibitem [{\citenamefont {Hentrich}\ \emph {et~al.}(2018)\citenamefont
  {Hentrich}, \citenamefont {Wolter}, \citenamefont {Zotos}, \citenamefont
  {Brenig}, \citenamefont {Nowak}, \citenamefont {Isaeva}, \citenamefont
  {Doert}, \citenamefont {Banerjee}, \citenamefont {Lampen-Kelley},
  \citenamefont {Mandrus}, \citenamefont {Nagler}, \citenamefont {Sears},
  \citenamefont {Kim}, \citenamefont {B\"uchner},\ and\ \citenamefont
  {Hess}}]{Hentrich2018}%
  \BibitemOpen
  \bibfield  {author} {\bibinfo {author} {\bibfnamefont {R.}~\bibnamefont
  {Hentrich}}, \bibinfo {author} {\bibfnamefont {A.~U.~B.}\ \bibnamefont
  {Wolter}}, \bibinfo {author} {\bibfnamefont {X.}~\bibnamefont {Zotos}},
  \bibinfo {author} {\bibfnamefont {W.}~\bibnamefont {Brenig}}, \bibinfo
  {author} {\bibfnamefont {D.}~\bibnamefont {Nowak}}, \bibinfo {author}
  {\bibfnamefont {A.}~\bibnamefont {Isaeva}}, \bibinfo {author} {\bibfnamefont
  {T.}~\bibnamefont {Doert}}, \bibinfo {author} {\bibfnamefont
  {A.}~\bibnamefont {Banerjee}}, \bibinfo {author} {\bibfnamefont
  {P.}~\bibnamefont {Lampen-Kelley}}, \bibinfo {author} {\bibfnamefont {D.~G.}\
  \bibnamefont {Mandrus}}, \bibinfo {author} {\bibfnamefont {S.~E.}\
  \bibnamefont {Nagler}}, \bibinfo {author} {\bibfnamefont {J.}~\bibnamefont
  {Sears}}, \bibinfo {author} {\bibfnamefont {Y.-J.}\ \bibnamefont {Kim}},
  \bibinfo {author} {\bibfnamefont {B.}~\bibnamefont {B\"uchner}}, \ and\
  \bibinfo {author} {\bibfnamefont {C.}~\bibnamefont {Hess}},\ }\href@noop {}
  {\bibfield  {journal} {\bibinfo  {journal} {Phys. Rev. Lett.}\ }\textbf
  {\bibinfo {volume} {120}},\ \bibinfo {pages} {117204} (\bibinfo {year}
  {2018})}\BibitemShut {NoStop}%
\bibitem [{\citenamefont {Pershoguba}\ \emph {et~al.}(2018)\citenamefont
  {Pershoguba}, \citenamefont {Banerjee}, \citenamefont {Lashley},
  \citenamefont {Park}, \citenamefont {\AA{}gren}, \citenamefont {Aeppli},\
  and\ \citenamefont {Balatsky}}]{Pershoguba2018}%
  \BibitemOpen
  \bibfield  {author} {\bibinfo {author} {\bibfnamefont {S.~S.}\ \bibnamefont
  {Pershoguba}}, \bibinfo {author} {\bibfnamefont {S.}~\bibnamefont
  {Banerjee}}, \bibinfo {author} {\bibfnamefont {J.~C.}\ \bibnamefont
  {Lashley}}, \bibinfo {author} {\bibfnamefont {J.}~\bibnamefont {Park}},
  \bibinfo {author} {\bibfnamefont {H.}~\bibnamefont {\AA{}gren}}, \bibinfo
  {author} {\bibfnamefont {G.}~\bibnamefont {Aeppli}}, \ and\ \bibinfo {author}
  {\bibfnamefont {A.~V.}\ \bibnamefont {Balatsky}},\ }\href@noop {} {\bibfield
  {journal} {\bibinfo  {journal} {Phys. Rev. X}\ }\textbf {\bibinfo {volume}
  {8}},\ \bibinfo {pages} {011010} (\bibinfo {year} {2018})}\BibitemShut
  {NoStop}%
\bibitem [{\citenamefont {McGuire}\ \emph {et~al.}(2015)\citenamefont
  {McGuire}, \citenamefont {Dixit}, \citenamefont {Cooper},\ and\ \citenamefont
  {Sales}}]{rmdcs}%
  \BibitemOpen
  \bibfield  {author} {\bibinfo {author} {\bibfnamefont {M.~A.}\ \bibnamefont
  {McGuire}}, \bibinfo {author} {\bibfnamefont {H.}~\bibnamefont {Dixit}},
  \bibinfo {author} {\bibfnamefont {V.~R.}\ \bibnamefont {Cooper}}, \ and\
  \bibinfo {author} {\bibfnamefont {B.~C.}\ \bibnamefont {Sales}},\ }\href@noop
  {} {\bibfield  {journal} {\bibinfo  {journal} {Chem. Mater.}\ }\textbf
  {\bibinfo {volume} {27}},\ \bibinfo {pages} {612} (\bibinfo {year}
  {2015})}\BibitemShut {NoStop}%
\bibitem [{\citenamefont {Huang}\ \emph {et~al.}(2017)\citenamefont {Huang},
  \citenamefont {Clark}, \citenamefont {Navarro-Moratalla}, \citenamefont
  {Klein}, \citenamefont {Cheng}, \citenamefont {Seyler}, \citenamefont {Ding},
  \citenamefont {Schmidgall}, \citenamefont {McGuire}, \citenamefont {Cobden},
  \citenamefont {Wang}, \citenamefont {Di}, \citenamefont {Jarillo-Herrero},\
  and\ \citenamefont {Xu}}]{Huang2017}%
  \BibitemOpen
  \bibfield  {author} {\bibinfo {author} {\bibfnamefont {B.}~\bibnamefont
  {Huang}}, \bibinfo {author} {\bibfnamefont {G.}~\bibnamefont {Clark}},
  \bibinfo {author} {\bibfnamefont {E.}~\bibnamefont {Navarro-Moratalla}},
  \bibinfo {author} {\bibfnamefont {D.~R.}\ \bibnamefont {Klein}}, \bibinfo
  {author} {\bibfnamefont {R.}~\bibnamefont {Cheng}}, \bibinfo {author}
  {\bibfnamefont {K.~L.}\ \bibnamefont {Seyler}}, \bibinfo {author}
  {\bibfnamefont {Z.}~\bibnamefont {Ding}}, \bibinfo {author} {\bibfnamefont
  {E.}~\bibnamefont {Schmidgall}}, \bibinfo {author} {\bibfnamefont {M.~A.}\
  \bibnamefont {McGuire}}, \bibinfo {author} {\bibfnamefont {D.~H.}\
  \bibnamefont {Cobden}}, \bibinfo {author} {\bibfnamefont {Y.}~\bibnamefont
  {Wang}}, \bibinfo {author} {\bibfnamefont {X.}~\bibnamefont {Di}}, \bibinfo
  {author} {\bibfnamefont {P.}~\bibnamefont {Jarillo-Herrero}}, \ and\ \bibinfo
  {author} {\bibfnamefont {X.}~\bibnamefont {Xu}},\ }\href@noop {} {\bibfield
  {journal} {\bibinfo  {journal} {Nature}\ }\textbf {\bibinfo {volume} {546}},\
  \bibinfo {pages} {270} (\bibinfo {year} {2017})}\BibitemShut {NoStop}%
\bibitem [{\citenamefont {Hansen}\ and\ \citenamefont
  {Griffel}(1958)}]{Hansen1958}%
  \BibitemOpen
  \bibfield  {author} {\bibinfo {author} {\bibfnamefont {W.~N.}\ \bibnamefont
  {Hansen}}\ and\ \bibinfo {author} {\bibfnamefont {M.}~\bibnamefont
  {Griffel}},\ }\href@noop {} {\bibfield  {journal} {\bibinfo  {journal} {J.
  Chem. Phys.}\ }\textbf {\bibinfo {volume} {28}},\ \bibinfo {pages} {902}
  (\bibinfo {year} {1958})}\BibitemShut {NoStop}%
\bibitem [{\citenamefont {Cable}\ \emph {et~al.}(1961)\citenamefont {Cable},
  \citenamefont {Wilkinson},\ and\ \citenamefont {Wollan}}]{Cable1961}%
  \BibitemOpen
  \bibfield  {author} {\bibinfo {author} {\bibfnamefont {J.~W.}\ \bibnamefont
  {Cable}}, \bibinfo {author} {\bibfnamefont {M.~K.}\ \bibnamefont
  {Wilkinson}}, \ and\ \bibinfo {author} {\bibfnamefont {E.~O.}\ \bibnamefont
  {Wollan}},\ }\href@noop {} {\bibfield  {journal} {\bibinfo  {journal} {J.
  Phys. Chem. Solids}\ }\textbf {\bibinfo {volume} {19}},\ \bibinfo {pages}
  {29} (\bibinfo {year} {1961})}\BibitemShut {NoStop}%
\bibitem [{\citenamefont {Bizette}\ \emph {et~al.}(1961)\citenamefont
  {Bizette}, \citenamefont {Terrier},\ and\ \citenamefont
  {Adam}}]{Bizette1961}%
  \BibitemOpen
  \bibfield  {author} {\bibinfo {author} {\bibfnamefont {H.}~\bibnamefont
  {Bizette}}, \bibinfo {author} {\bibfnamefont {C.}~\bibnamefont {Terrier}}, \
  and\ \bibinfo {author} {\bibfnamefont {A.}~\bibnamefont {Adam}},\ }\href@noop
  {} {\bibfield  {journal} {\bibinfo  {journal} {C. R. Acad. Sci.}\ }\textbf
  {\bibinfo {volume} {252}},\ \bibinfo {pages} {1571} (\bibinfo {year}
  {1961})}\BibitemShut {NoStop}%
\bibitem [{\citenamefont {Narath}\ and\ \citenamefont
  {Davis}(1965)}]{Narath1965}%
  \BibitemOpen
  \bibfield  {author} {\bibinfo {author} {\bibfnamefont {A.}~\bibnamefont
  {Narath}}\ and\ \bibinfo {author} {\bibfnamefont {H.~L.}\ \bibnamefont
  {Davis}},\ }\href@noop {} {\bibfield  {journal} {\bibinfo  {journal} {Phys.
  Rev.}\ }\textbf {\bibinfo {volume} {137}},\ \bibinfo {pages} {A163} (\bibinfo
  {year} {1965})}\BibitemShut {NoStop}%
\bibitem [{\citenamefont {Kuhlow}(1982)}]{Kuhlow1982}%
  \BibitemOpen
  \bibfield  {author} {\bibinfo {author} {\bibfnamefont {B.}~\bibnamefont
  {Kuhlow}},\ }\href@noop {} {\bibfield  {journal} {\bibinfo  {journal} {Phys.
  Stat. Sol.}\ }\textbf {\bibinfo {volume} {72}},\ \bibinfo {pages} {161}
  (\bibinfo {year} {1982})}\BibitemShut {NoStop}%
\bibitem [{\citenamefont {Baibich}\ \emph {et~al.}(1988)\citenamefont
  {Baibich}, \citenamefont {Broto}, \citenamefont {Fert}, \citenamefont {Dau},
  \citenamefont {Petroff}, \citenamefont {Etienne}, \citenamefont {Creuzet},
  \citenamefont {Friederich},\ and\ \citenamefont {Chazelas}}]{rgmr}%
  \BibitemOpen
  \bibfield  {author} {\bibinfo {author} {\bibfnamefont {M.~N.}\ \bibnamefont
  {Baibich}}, \bibinfo {author} {\bibfnamefont {J.~M.}\ \bibnamefont {Broto}},
  \bibinfo {author} {\bibfnamefont {A.}~\bibnamefont {Fert}}, \bibinfo {author}
  {\bibfnamefont {F.~N.~V.}\ \bibnamefont {Dau}}, \bibinfo {author}
  {\bibfnamefont {F.}~\bibnamefont {Petroff}}, \bibinfo {author} {\bibfnamefont
  {P.}~\bibnamefont {Etienne}}, \bibinfo {author} {\bibfnamefont
  {G.}~\bibnamefont {Creuzet}}, \bibinfo {author} {\bibfnamefont
  {A.}~\bibnamefont {Friederich}}, \ and\ \bibinfo {author} {\bibfnamefont
  {J.}~\bibnamefont {Chazelas}},\ }\href@noop {} {\bibfield  {journal}
  {\bibinfo  {journal} {Phys. Rev. Lett.}\ }\textbf {\bibinfo {volume} {61}},\
  \bibinfo {pages} {2472} (\bibinfo {year} {1988})}\BibitemShut {NoStop}%
\bibitem [{\citenamefont {Callaway}(1959)}]{Callaway1959}%
  \BibitemOpen
  \bibfield  {author} {\bibinfo {author} {\bibfnamefont {J.}~\bibnamefont
  {Callaway}},\ }\href@noop {} {\bibfield  {journal} {\bibinfo  {journal}
  {Phys. Rev.}\ }\textbf {\bibinfo {volume} {113}},\ \bibinfo {pages} {1046}
  (\bibinfo {year} {1959})}\BibitemShut {NoStop}%
\bibitem [{\citenamefont {Bergman}(1979)}]{Bergmanbook}%
  \BibitemOpen
  \bibfield  {author} {\bibinfo {author} {\bibfnamefont {R.}~\bibnamefont
  {Bergman}},\ }\href@noop {} {\emph {\bibinfo {title} {Thermal conduction in
  solids}}}\ (\bibinfo  {publisher} {Oxford University Press},\ \bibinfo
  {address} {Oxford},\ \bibinfo {year} {1979})\BibitemShut {NoStop}%
\bibitem [{\citenamefont {Sologubenko}\ \emph {et~al.}(2001)\citenamefont
  {Sologubenko}, \citenamefont {Giann\`o}, \citenamefont {Ott}, \citenamefont
  {Vietkine},\ and\ \citenamefont {Revcolevschi}}]{Sologubenko2001}%
  \BibitemOpen
  \bibfield  {author} {\bibinfo {author} {\bibfnamefont {A.~V.}\ \bibnamefont
  {Sologubenko}}, \bibinfo {author} {\bibfnamefont {K.}~\bibnamefont
  {Giann\`o}}, \bibinfo {author} {\bibfnamefont {H.~R.}\ \bibnamefont {Ott}},
  \bibinfo {author} {\bibfnamefont {A.}~\bibnamefont {Vietkine}}, \ and\
  \bibinfo {author} {\bibfnamefont {A.}~\bibnamefont {Revcolevschi}},\
  }\href@noop {} {\bibfield  {journal} {\bibinfo  {journal} {Phys. Rev. B}\
  }\textbf {\bibinfo {volume} {64}},\ \bibinfo {pages} {054412} (\bibinfo
  {year} {2001})}\BibitemShut {NoStop}%
\bibitem [{\citenamefont {Glamazda}\ \emph {et~al.}(2017)\citenamefont
  {Glamazda}, \citenamefont {Lemmens}, \citenamefont {Do}, \citenamefont
  {Kwon},\ and\ \citenamefont {Choi}}]{Glamazda2017}%
  \BibitemOpen
  \bibfield  {author} {\bibinfo {author} {\bibfnamefont {A.}~\bibnamefont
  {Glamazda}}, \bibinfo {author} {\bibfnamefont {P.}~\bibnamefont {Lemmens}},
  \bibinfo {author} {\bibfnamefont {S.-H.}\ \bibnamefont {Do}}, \bibinfo
  {author} {\bibfnamefont {Y.~S.}\ \bibnamefont {Kwon}}, \ and\ \bibinfo
  {author} {\bibfnamefont {K.-Y.}\ \bibnamefont {Choi}},\ }\href@noop {}
  {\bibfield  {journal} {\bibinfo  {journal} {Phys. Rev. B}\ }\textbf {\bibinfo
  {volume} {95}},\ \bibinfo {pages} {174429} (\bibinfo {year}
  {2017})}\BibitemShut {NoStop}%
\bibitem [{\citenamefont {McGuire}\ \emph {et~al.}(2017)\citenamefont
  {McGuire}, \citenamefont {Clark}, \citenamefont {KC}, \citenamefont {Chance},
  \citenamefont {Jellison}, \citenamefont {Cooper}, \citenamefont {Xu},\ and\
  \citenamefont {Sales}}]{McGuire2017}%
  \BibitemOpen
  \bibfield  {author} {\bibinfo {author} {\bibfnamefont {M.~A.}\ \bibnamefont
  {McGuire}}, \bibinfo {author} {\bibfnamefont {G.}~\bibnamefont {Clark}},
  \bibinfo {author} {\bibfnamefont {S.}~\bibnamefont {KC}}, \bibinfo {author}
  {\bibfnamefont {W.~M.}\ \bibnamefont {Chance}}, \bibinfo {author}
  {\bibfnamefont {G.~E.}\ \bibnamefont {Jellison}}, \bibinfo {author}
  {\bibfnamefont {V.~R.}\ \bibnamefont {Cooper}}, \bibinfo {author}
  {\bibfnamefont {X.}~\bibnamefont {Xu}}, \ and\ \bibinfo {author}
  {\bibfnamefont {B.~C.}\ \bibnamefont {Sales}},\ }\href@noop {} {\bibfield
  {journal} {\bibinfo  {journal} {Phys. Rev. Mater.}\ }\textbf {\bibinfo
  {volume} {1}},\ \bibinfo {pages} {014001} (\bibinfo {year}
  {2017})}\BibitemShut {NoStop}%
\bibitem [{\citenamefont {Morosin}\ and\ \citenamefont
  {Narath}(1964)}]{Morosin1964}%
  \BibitemOpen
  \bibfield  {author} {\bibinfo {author} {\bibfnamefont {B.}~\bibnamefont
  {Morosin}}\ and\ \bibinfo {author} {\bibfnamefont {A.}~\bibnamefont
  {Narath}},\ }\href@noop {} {\bibfield  {journal} {\bibinfo  {journal} {J.
  Chem. Phys.}\ }\textbf {\bibinfo {volume} {40}},\ \bibinfo {pages} {1958}
  (\bibinfo {year} {1964})}\BibitemShut {NoStop}%
\bibitem [{\citenamefont {Pohl}\ and\ \citenamefont
  {Stritzker}(1982)}]{Pohl1982}%
  \BibitemOpen
  \bibfield  {author} {\bibinfo {author} {\bibfnamefont {R.~O.}\ \bibnamefont
  {Pohl}}\ and\ \bibinfo {author} {\bibfnamefont {B.}~\bibnamefont
  {Stritzker}},\ }\href@noop {} {\bibfield  {journal} {\bibinfo  {journal}
  {Phys. Rev. B}\ }\textbf {\bibinfo {volume} {25}},\ \bibinfo {pages} {3068}
  (\bibinfo {year} {1982})}\BibitemShut {NoStop}%
\bibitem [{\citenamefont {Do}\ \emph {et~al.}(2017)\citenamefont {Do},
  \citenamefont {Park}, \citenamefont {Yoshitake}, \citenamefont {Nasu},
  \citenamefont {Motome}, \citenamefont {Kwon}, \citenamefont {Adroja},
  \citenamefont {Voneshen}, \citenamefont {Kim}, \citenamefont {Jang},
  \citenamefont {Park}, \citenamefont {Choi},\ and\ \citenamefont
  {Ji}}]{HwanDo2017}%
  \BibitemOpen
  \bibfield  {author} {\bibinfo {author} {\bibfnamefont {S.-H.}\ \bibnamefont
  {Do}}, \bibinfo {author} {\bibfnamefont {S.-Y.}\ \bibnamefont {Park}},
  \bibinfo {author} {\bibfnamefont {J.}~\bibnamefont {Yoshitake}}, \bibinfo
  {author} {\bibfnamefont {J.}~\bibnamefont {Nasu}}, \bibinfo {author}
  {\bibfnamefont {Y.}~\bibnamefont {Motome}}, \bibinfo {author} {\bibfnamefont
  {Y.~S.}\ \bibnamefont {Kwon}}, \bibinfo {author} {\bibfnamefont {D.~T.}\
  \bibnamefont {Adroja}}, \bibinfo {author} {\bibfnamefont {D.~J.}\
  \bibnamefont {Voneshen}}, \bibinfo {author} {\bibfnamefont {K.}~\bibnamefont
  {Kim}}, \bibinfo {author} {\bibfnamefont {T.-H.}\ \bibnamefont {Jang}},
  \bibinfo {author} {\bibfnamefont {J.-H.}\ \bibnamefont {Park}}, \bibinfo
  {author} {\bibfnamefont {K.-Y.}\ \bibnamefont {Choi}}, \ and\ \bibinfo
  {author} {\bibfnamefont {S.}~\bibnamefont {Ji}},\ }\href@noop {} {\bibfield
  {journal} {\bibinfo  {journal} {Nature Phys.}\ }\textbf {\bibinfo {volume}
  {13}},\ \bibinfo {pages} {1079} (\bibinfo {year} {2017})}\BibitemShut
  {NoStop}%
\bibitem [{\citenamefont {Hirschberger}\ \emph {et~al.}(2015)\citenamefont
  {Hirschberger}, \citenamefont {Chisnell}, \citenamefont {Lee},\ and\
  \citenamefont {Ong}}]{Hirschberger2015}%
  \BibitemOpen
  \bibfield  {author} {\bibinfo {author} {\bibfnamefont {M.}~\bibnamefont
  {Hirschberger}}, \bibinfo {author} {\bibfnamefont {R.}~\bibnamefont
  {Chisnell}}, \bibinfo {author} {\bibfnamefont {Y.~S.}\ \bibnamefont {Lee}}, \
  and\ \bibinfo {author} {\bibfnamefont {N.~P.}\ \bibnamefont {Ong}},\
  }\href@noop {} {\bibfield  {journal} {\bibinfo  {journal} {Phys. Rev. Lett.}\
  }\textbf {\bibinfo {volume} {115}},\ \bibinfo {pages} {106603} (\bibinfo
  {year} {2015})}\BibitemShut {NoStop}%
\bibitem [{\citenamefont {Pan}\ \emph {et~al.}(2013)\citenamefont {Pan},
  \citenamefont {Guan}, \citenamefont {Hong}, \citenamefont {Zhou},
  \citenamefont {Qiu}, \citenamefont {Zhang},\ and\ \citenamefont
  {Li}}]{Pan2013}%
  \BibitemOpen
  \bibfield  {author} {\bibinfo {author} {\bibfnamefont {B.~Y.}\ \bibnamefont
  {Pan}}, \bibinfo {author} {\bibfnamefont {T.~Y.}\ \bibnamefont {Guan}},
  \bibinfo {author} {\bibfnamefont {X.~C.}\ \bibnamefont {Hong}}, \bibinfo
  {author} {\bibfnamefont {S.~Y.}\ \bibnamefont {Zhou}}, \bibinfo {author}
  {\bibfnamefont {X.}~\bibnamefont {Qiu}}, \bibinfo {author} {\bibfnamefont
  {H.}~\bibnamefont {Zhang}}, \ and\ \bibinfo {author} {\bibfnamefont {S.~Y.}\
  \bibnamefont {Li}},\ }\href@noop {} {\bibfield  {journal} {\bibinfo
  {journal} {Europhys. Lett.}\ }\textbf {\bibinfo {volume} {103}},\ \bibinfo
  {pages} {37005} (\bibinfo {year} {2013})}\BibitemShut {NoStop}%
\bibitem [{\citenamefont {Blundell}(2001)}]{Blundellbook}%
  \BibitemOpen
  \bibfield  {author} {\bibinfo {author} {\bibfnamefont {S.}~\bibnamefont
  {Blundell}},\ }\href@noop {} {\emph {\bibinfo {title} {Magnetism in Condensed
  Matter}}}\ (\bibinfo  {publisher} {Oxford University Press, Oxford},\
  \bibinfo {year} {2001})\BibitemShut {NoStop}%
\bibitem [{\citenamefont {Klein}\ \emph {et~al.}(2019)\citenamefont {Klein},
  \citenamefont {MacNeill}, \citenamefont {Song}, \citenamefont {Larson},
  \citenamefont {Fang}, \citenamefont {Xu}, \citenamefont {Ribeiro},
  \citenamefont {Canfield}, \citenamefont {Kaxiras}, \citenamefont {Comin},\
  and\ \citenamefont {Jarillo-Herrero}}]{kleinetal}%
  \BibitemOpen
  \bibfield  {author} {\bibinfo {author} {\bibfnamefont {D.~R.}\ \bibnamefont
  {Klein}}, \bibinfo {author} {\bibfnamefont {D.}~\bibnamefont {MacNeill}},
  \bibinfo {author} {\bibfnamefont {Q.}~\bibnamefont {Song}}, \bibinfo {author}
  {\bibfnamefont {D.~T.}\ \bibnamefont {Larson}}, \bibinfo {author}
  {\bibfnamefont {S.}~\bibnamefont {Fang}}, \bibinfo {author} {\bibfnamefont
  {M.}~\bibnamefont {Xu}}, \bibinfo {author} {\bibfnamefont {R.~A.}\
  \bibnamefont {Ribeiro}}, \bibinfo {author} {\bibfnamefont {P.~C.}\
  \bibnamefont {Canfield}}, \bibinfo {author} {\bibfnamefont {E.}~\bibnamefont
  {Kaxiras}}, \bibinfo {author} {\bibfnamefont {R.}~\bibnamefont {Comin}}, \
  and\ \bibinfo {author} {\bibfnamefont {P.}~\bibnamefont {Jarillo-Herrero}},\
  }\href@noop {} {\bibfield  {journal} {\bibinfo  {journal} {Nature Phys.}\
  }\textbf {\bibinfo {volume} {15}},\ \bibinfo {pages} {in press;
  arXiv:1903.00002} (\bibinfo {year} {2019})}\BibitemShut {NoStop}%
\bibitem [{\citenamefont {Kitaev}(2006)}]{Kitaev2006}%
  \BibitemOpen
  \bibfield  {author} {\bibinfo {author} {\bibfnamefont {A.}~\bibnamefont
  {Kitaev}},\ }\href@noop {} {\bibfield  {journal} {\bibinfo  {journal} {Ann.
  Phys.}\ }\textbf {\bibinfo {volume} {321}},\ \bibinfo {pages} {2} (\bibinfo
  {year} {2006})}\BibitemShut {NoStop}%
\bibitem [{\citenamefont {Wang}\ \emph {et~al.}()\citenamefont {Wang},
  \citenamefont {Normand},\ and\ \citenamefont {Liu}}]{rwnl}%
  \BibitemOpen
  \bibfield  {author} {\bibinfo {author} {\bibfnamefont {J.}~\bibnamefont
  {Wang}}, \bibinfo {author} {\bibfnamefont {B.}~\bibnamefont {Normand}}, \
  and\ \bibinfo {author} {\bibfnamefont {Z.-X.}\ \bibnamefont {Liu}},\
  }\href@noop {} {\bibinfo  {journal} {to appear in Phys. Rev. Lett.
  (arXiv:1903.10026)}\ }\BibitemShut {NoStop}%
\bibitem [{\citenamefont {Liu}\ and\ \citenamefont {Normand}(2018)}]{rln}%
  \BibitemOpen
\bibfield  {journal} {  }\bibfield  {author} {\bibinfo {author} {\bibfnamefont
  {Z.-X.}\ \bibnamefont {Liu}}\ and\ \bibinfo {author} {\bibfnamefont
  {B.}~\bibnamefont {Normand}},\ }\href@noop {} {\bibfield  {journal} {\bibinfo
   {journal} {Phys. Rev. Lett.}\ }\textbf {\bibinfo {volume} {120}},\ \bibinfo
  {pages} {187201} (\bibinfo {year} {2018})}\BibitemShut {NoStop}%
\bibitem [{\citenamefont {Ye}\ \emph {et~al.}(2018)\citenamefont {Ye},
  \citenamefont {Halasz}, \citenamefont {Savary},\ and\ \citenamefont
  {Balents}}]{ryhsb}%
  \BibitemOpen
  \bibfield  {author} {\bibinfo {author} {\bibfnamefont {M.}~\bibnamefont
  {Ye}}, \bibinfo {author} {\bibfnamefont {G.~B.}\ \bibnamefont {Halasz}},
  \bibinfo {author} {\bibfnamefont {L.}~\bibnamefont {Savary}}, \ and\ \bibinfo
  {author} {\bibfnamefont {L.}~\bibnamefont {Balents}},\ }\href@noop {}
  {\bibfield  {journal} {\bibinfo  {journal} {Phys. Rev. Lett.}\ }\textbf
  {\bibinfo {volume} {121}},\ \bibinfo {pages} {147201} (\bibinfo {year}
  {2018})}\BibitemShut {NoStop}%
\bibitem [{\citenamefont {Vinkler-Aviv}\ and\ \citenamefont
  {Rosch}(2018)}]{rvr}%
  \BibitemOpen
  \bibfield  {author} {\bibinfo {author} {\bibfnamefont {Y.}~\bibnamefont
  {Vinkler-Aviv}}\ and\ \bibinfo {author} {\bibfnamefont {A.}~\bibnamefont
  {Rosch}},\ }\href@noop {} {\bibfield  {journal} {\bibinfo  {journal} {Phys.
  Rev. X}\ }\textbf {\bibinfo {volume} {8}},\ \bibinfo {pages} {031032}
  (\bibinfo {year} {2018})}\BibitemShut {NoStop}%
\bibitem [{\citenamefont {Kasahara}\ \emph {et~al.}(2018)\citenamefont
  {Kasahara}, \citenamefont {Ohnishi}, \citenamefont {Mizukami}, \citenamefont
  {Tanaka}, \citenamefont {Ma}, \citenamefont {Sugii}, \citenamefont {Kurita},
  \citenamefont {Tanaka}, \citenamefont {Nasu}, \citenamefont {Motome},
  \citenamefont {Shibauchi},\ and\ \citenamefont {Matsuda}}]{rkbs}%
  \BibitemOpen
  \bibfield  {author} {\bibinfo {author} {\bibfnamefont {Y.}~\bibnamefont
  {Kasahara}}, \bibinfo {author} {\bibfnamefont {T.}~\bibnamefont {Ohnishi}},
  \bibinfo {author} {\bibfnamefont {Y.}~\bibnamefont {Mizukami}}, \bibinfo
  {author} {\bibfnamefont {O.}~\bibnamefont {Tanaka}}, \bibinfo {author}
  {\bibfnamefont {S.}~\bibnamefont {Ma}}, \bibinfo {author} {\bibfnamefont
  {K.}~\bibnamefont {Sugii}}, \bibinfo {author} {\bibfnamefont
  {N.}~\bibnamefont {Kurita}}, \bibinfo {author} {\bibfnamefont
  {H.}~\bibnamefont {Tanaka}}, \bibinfo {author} {\bibfnamefont
  {J.}~\bibnamefont {Nasu}}, \bibinfo {author} {\bibfnamefont {Y.}~\bibnamefont
  {Motome}}, \bibinfo {author} {\bibfnamefont {T.}~\bibnamefont {Shibauchi}}, \
  and\ \bibinfo {author} {\bibfnamefont {Y.}~\bibnamefont {Matsuda}},\
  }\href@noop {} {\bibfield  {journal} {\bibinfo  {journal} {Nature}\ }\textbf
  {\bibinfo {volume} {559}},\ \bibinfo {pages} {227} (\bibinfo {year}
  {2018})}\BibitemShut {NoStop}%
\bibitem [{\citenamefont {Cai}\ \emph {et~al.}(2019)\citenamefont {Cai},
  \citenamefont {Song}, \citenamefont {Wilson}, \citenamefont {Clark},
  \citenamefont {He}, \citenamefont {Zhang}, \citenamefont {Taniguchi},
  \citenamefont {Watanabe}, \citenamefont {Yao}, \citenamefont {Xiao},
  \citenamefont {McGuire}, \citenamefont {Cobden},\ and\ \citenamefont
  {Xu}}]{caoetal}%
  \BibitemOpen
  \bibfield  {author} {\bibinfo {author} {\bibfnamefont {X.}~\bibnamefont
  {Cai}}, \bibinfo {author} {\bibfnamefont {T.}~\bibnamefont {Song}}, \bibinfo
  {author} {\bibfnamefont {N.~P.}\ \bibnamefont {Wilson}}, \bibinfo {author}
  {\bibfnamefont {G.}~\bibnamefont {Clark}}, \bibinfo {author} {\bibfnamefont
  {M.}~\bibnamefont {He}}, \bibinfo {author} {\bibfnamefont {X.}~\bibnamefont
  {Zhang}}, \bibinfo {author} {\bibfnamefont {T.}~\bibnamefont {Taniguchi}},
  \bibinfo {author} {\bibfnamefont {K.}~\bibnamefont {Watanabe}}, \bibinfo
  {author} {\bibfnamefont {W.}~\bibnamefont {Yao}}, \bibinfo {author}
  {\bibfnamefont {D.}~\bibnamefont {Xiao}}, \bibinfo {author} {\bibfnamefont
  {M.~A.}\ \bibnamefont {McGuire}}, \bibinfo {author} {\bibfnamefont {D.~H.}\
  \bibnamefont {Cobden}}, \ and\ \bibinfo {author} {\bibfnamefont
  {X.}~\bibnamefont {Xu}},\ }\href@noop {} {\bibfield  {journal} {\bibinfo
  {journal} {Nano Lett.}\ }\textbf {\bibinfo {volume} {19}},\ \bibinfo {pages}
  {3993} (\bibinfo {year} {2019})}\BibitemShut {NoStop}%
\bibitem [{\citenamefont {Kim}\ \emph {et~al.}(2019)\citenamefont {Kim},
  \citenamefont {Yang}, \citenamefont {Li}, \citenamefont {Jiang},
  \citenamefont {Jin}, \citenamefont {Tao}, \citenamefont {Nichols},
  \citenamefont {Sfigakis}, \citenamefont {Zhong}, \citenamefont {Li},
  \citenamefont {Tian}, \citenamefont {Cory}, \citenamefont {Miao},
  \citenamefont {Shan}, \citenamefont {Mak}, \citenamefont {Lei}, \citenamefont
  {Sun}, \citenamefont {Zhao},\ and\ \citenamefont {Tsen}}]{kimetal}%
  \BibitemOpen
  \bibfield  {author} {\bibinfo {author} {\bibfnamefont {H.~H.}\ \bibnamefont
  {Kim}}, \bibinfo {author} {\bibfnamefont {B.}~\bibnamefont {Yang}}, \bibinfo
  {author} {\bibfnamefont {S.}~\bibnamefont {Li}}, \bibinfo {author}
  {\bibfnamefont {S.}~\bibnamefont {Jiang}}, \bibinfo {author} {\bibfnamefont
  {C.}~\bibnamefont {Jin}}, \bibinfo {author} {\bibfnamefont {Z.}~\bibnamefont
  {Tao}}, \bibinfo {author} {\bibfnamefont {G.}~\bibnamefont {Nichols}},
  \bibinfo {author} {\bibfnamefont {F.}~\bibnamefont {Sfigakis}}, \bibinfo
  {author} {\bibfnamefont {S.}~\bibnamefont {Zhong}}, \bibinfo {author}
  {\bibfnamefont {C.}~\bibnamefont {Li}}, \bibinfo {author} {\bibfnamefont
  {S.}~\bibnamefont {Tian}}, \bibinfo {author} {\bibfnamefont {D.~G.}\
  \bibnamefont {Cory}}, \bibinfo {author} {\bibfnamefont {G.-X.}\ \bibnamefont
  {Miao}}, \bibinfo {author} {\bibfnamefont {J.}~\bibnamefont {Shan}}, \bibinfo
  {author} {\bibfnamefont {K.~F.}\ \bibnamefont {Mak}}, \bibinfo {author}
  {\bibfnamefont {H.}~\bibnamefont {Lei}}, \bibinfo {author} {\bibfnamefont
  {K.}~\bibnamefont {Sun}}, \bibinfo {author} {\bibfnamefont {L.}~\bibnamefont
  {Zhao}}, \ and\ \bibinfo {author} {\bibfnamefont {A.~W.}\ \bibnamefont
  {Tsen}},\ }\href@noop {} {\bibfield  {journal} {\bibinfo  {journal} {Proc.
  Natl. Acad. Sci. U.S.A.}\ }\textbf {\bibinfo {volume} {116}},\ \bibinfo
  {pages} {11131} (\bibinfo {year} {2019})}\BibitemShut {NoStop}%
\end{thebibliography}

\end{document}